\documentclass[aps,prx,twocolumn]{revtex4-1}
\usepackage{amscd,amssymb,amsmath,latexsym,bm}
\usepackage[mathcal,mathscr]{euscript}
\usepackage{lipsum}
\usepackage{amsfonts}
\usepackage{amsmath}
\usepackage{amssymb}
\usepackage{txfonts}
\usepackage{pxfonts}
\usepackage{graphicx}
\usepackage{bm,units,yfonts}
\usepackage[table,dvipsnames]{xcolor}
\usepackage{hyperref}
\usepackage{lineno}

\newcommand{\CM}{{\mathbb C}}

\newcommand{\NM}{{\mathbb N}}

\newcommand{\RM}{{\mathbb R}}
\newcommand{\SM}{{\mathbb S}}

\newcommand{\ZM}{{\mathbb Z}}

\newcommand{\Aa}{{\mathcal A}}

\newcommand{\Ff}{{\mathcal F}}

\newcommand{\Hh}{{\mathcal H}}

\newcommand{\Tt}{{\mathcal T}}

\newcommand{\Mm}{{\mathcal M}}

\newcommand{\Ll}{{\mathcal L}}

\begin{document}

\title{Topological Gaps in Quasi-Periodic Spin Chains: A Numerical and K-Theoretic Analysis}

\author{Yifei Liu, Lea F. Santos, Emil Prodan}

\affiliation{
Department of Physics, Yeshiva University, New York, NY 10016, USA
}

\begin{abstract}
Topological phases supported by quasi-periodic spin-chain models and their bulk-boundary principles are investigated by numerical and K-theoretic methods. We show that, for both the un-correlated and correlated phases, the operator algebras that generate the Hamiltonians are non-commutative tori, hence the quasi-periodic chains display physics akin to the quantum Hall effect in two and higher dimensions. The robust topological edge modes are found to be strongly shaped by the interaction and, generically, they have hybrid edge-localized and chain-delocalized structures. Our findings lay the foundations for topological spin pumping using the phason of a quasi-periodic pattern as an adiabatic parameter, where selectively chosen quantized bits of magnetization can be transferred from one edge of the chain to the other. 
\end{abstract}

 \maketitle
 



\section{Introduction}

Engineering topological  gaps that host robust edge modes based on aperiodic principles is an extremely active area of research, spreading over several research fields such as condensed matter~\cite{YoshidaPRB2013,HeEPL2015,ProdanPRB2015,
TranPRB2015,VidalPRB2016,FulgaPRL2016,
CollinsNature2017,AgarwalaPRL2017,HuangPRL2018,BournJPA2018,
VarjasPRL2019,DevakulPRB2019,PaiPRB2019,KellendonkAHP2019,
ChenArxiv2019,IliasovPRB2020,FremlingPRB2020,
HuangPRB2020,ChenPRL2020,DuncanPRB2020}, photonics~\cite{KrausPRL2012,VerbinPRL2013,Vardeny2013,
TanesePRL2014,VerbinPRB2015,HuPRX2015,BandresPRX2016,
DareauPRL2017,BabouxPRB2017,ZilberbergNature2018,
KollarNature2019,KollarCMP2020,CarusottoNatPhys2020,
SchultheissAPX2020, YangLSA2020,ZhouLSA2020}, acoustics~\cite{ApigoPRL2019,NiCP2019,ChengPre2020} and mechanics~\cite{MitchellNature2018,MartinezPTRSA2018,ApigoPRM2018,RosaPRL2019,PalNJP2019, ZhouPRX2019,XiaPRAppl2020,RivaPRB2020,RivaPRB2020b,
XiaArxiv2020,RosaArxiv2020}. The basic working principles for all these studies rest on the existence of an intrinsic degree of freedom, the phason of the aperiodic pattern, which in many instances is experimentally accessible and, as such, it can be used as an adiabatic parameter in practical applications. For example, the first experimental demonstration of un-assisted dynamical edge-to-edge Thouless pumping \cite{ChengPre2020} has been achieved with such principles. In general, the phason space augments the physical space and supplies additional virtual dimensions \cite{ProdanPRB2015}, hence enabling physical phenomena beyond what can be ordinarily observed in our physical space. In particular, it can enable two and higher dimensional quantum Hall physics without the need of breaking the time-reversal and this has spurred the vigorous experimental progress mentioned above on the investigation of topological un-correlated phases from class A of classification table \cite{SRFL2008,QiPRB2008,Kit2009,RSFL2010}.

\vspace{0.2cm}

 An interesting and important ongoing research is characterizing the interplay between the aperiodicity, many-body correlations and topology in quantum systems or between aperiodicity, non-linear effects and topology in classical systems \cite{BarelliPRL1996, HePRA2013,HuPRB2016,ZengPRB2016,LiEPL2017,KunoNJP2017,
 MarraEPJ2017,TaddiaPRL2017,KePRA2017,NakagawaPRB2018,
 SarkarSR2018,HuPRA2019,LadoPRR2019,OritoPRB2019,
 ZuoNJP2020,ChenPRA2020,RosnerArxiv2020}. Quantum spin chains have been successfully used  in the past to shed some light on this question, especially because they can be simulated with modest computational resources. In particular, a quasi-periodic spin-system with tunned first and second nearest neighbor interactions has been used in \cite{HuPRB2016} to stabilized a fractional quantum Hall state. In \cite{LadoPRR2019}, it was shown that some correlated topological phases emerged under aperiodicity can be adiabatically connected to un-correlated phases, hence proving certain topological stability against correlations. The stability of the topological phases against many-body disorder was investigated in \cite{OritoPRB2019}. The main tool deployed in all these works is the first Chern number. However, as we shall see, quasi-periodic spin-chain systems host a plethora of higher Chern topological phases.

\vspace{0.2cm} 

Operator algebras and their K-theories emerged as natural frameworks for analyzing aperiodic systems \cite{Bellissard1986,Bellissard1995,KellendonkRMP95} and these are the tools we adopt in our study. The first task of such general program consists in the identification of the algebra that generates the quantum Hamiltonians. If this is successfully completed, then the K-theory of this algebra classifies the spectral projections of the Hamiltonians into classes that are invariant to continuous deformations of the models. In particular, every single spectral gap of the Hamiltonian receives a set of K-theoretic labels, which represent all topological invariants, both strong and weak, that can be associated to a gap. As it was pointed out in \cite{ProdanJGP2019}, this K-theoretic labels can be read off from certain maps of the integrated density of states (IDS). Such maps can be used to confirm that the algebra of the Hamiltonians was computed correctly or, in the cases when the algebras are unknown, the IDS maps can give hints on what the algebra might be. This general program has been carried out for several important classes of aperiodic un-correlated systems such as quasi-periodic \cite{ApigoPRM2018}, quasi-crystalline \cite{KellendonkAHP2019}, incommensurate \cite{ProdanJGP2019} and twisted \cite{RosaArxiv2020} bilayers. The progress with the correlated systems has been, however, very slow.
 
\vspace{0.2cm}

Working with quasi-periodic spin chains, where the $z$-component of the magnetization is conserved by the dynamics, we demonstrate first that the un-correlated Hamiltonians restricted to a magnetization sector $M=d$, where $d=1,2,3,\ldots$, belong to the non-commutative $d$-torus. This is confirmed by the numerically computed IDS maps and by the counts of the edge modes which conform with the bulk-boundary correspondence principles for these algebras. When a nearest neighbor interaction potential is turned on, in the regime of strong interaction, we observe a separation of the bulk energy spectrum in $d$ spectral islands \cite{JoelAJP2013} and, for $d$ up to three, we compute numerically the corresponding IDS maps. Surprisingly, every single feature seen in these maps can be explained by non-commutative tori. To explain the origins of these findings, we compute explicitly the generators of some of these algebras in the presence of strong interaction, which reveal the strongly correlated nature of the new topological phases. The topological edge modes are also found to be strongly shaped by the interaction. The analysis represents an important example where K-theory combined with numerical simulations are used to produce an extremely refined and complete picture of the topological phases supported by strongly interacting models and to establish quantitative bulk-boundary correspondence principles.

\vspace{0.2cm}

The paper is organized as it follows. In Sec.~\ref{Sec:AperiodicSpinChains}, we introduce the aperiodic pattern which is populated by $\frac{1}{2}$-spins and discuss the associated phason and phason space. In particular, we explain the special and general mechanism that allows one to use the phason as an adiabatic parameter and generate topological Thouless pumps. Also in Sec.~\ref{Sec:AperiodicSpinChains}, we describe the quantum spin models studied in our work. As we already mentioned, the non-commutative tori will play a central role in our analysis and, for this reason, we dedicate Sec.~\ref{Sec:NCT} to a review of these operators algebras and their K-theories. We discuss the K-theoretic gap labels, their relations to the standard topological invariants, their quantized range and how to compute them from the IDS maps. Sections~\ref{Sec:TopoGaps1} and \ref{Sec:TopoGaps1} are dedicated to the topological analysis of the un-correlated and strongly correlated spin chains, respectively. The last section summarizes the main conclusions of our work.

\section{Aperiodic Spin Chains}
\label{Sec:AperiodicSpinChains}

Any aperiodic pattern has an intrinsic degree of freedom, the phason, which lives on a smooth manifold when the pattern is quasiperiodic. Our spin chain models are defined over quasi-periodic patterns where the phason lives on a circle. In this section, we describe these patterns as well as the quantum spin models defined over them. Special attention is paid to the covariant property of the spin Hamiltonians and its implications.

\subsection{The aperiodic lattice}
\label{SSec:AperLattice}

\begin{figure}[t]
\includegraphics[width=\linewidth]{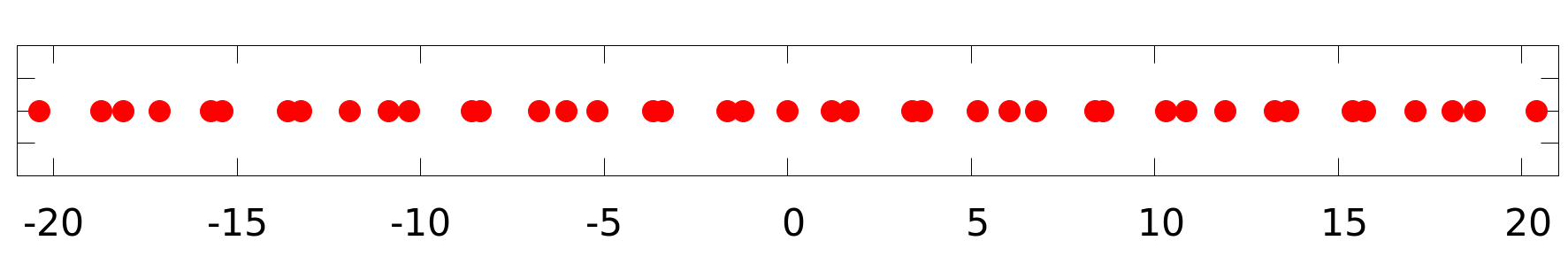}
\caption{\small A sample of a pattern generated with the algorithm \eqref{Eq:PattAlg}, using $r=0.45$, $\theta=\sqrt{2}$ and $\varphi=0$. The dots were given a finite size for visualization and some appear as overlapping, but this is not the actual case for the point pattern.}
\label{Fig:Pattern}
\end{figure}

We consider a spin-$\frac{1}{2}$ chain over a 1-dimensional lattice of points:
\begin{equation}
\Ll = \{p_n\}_{n = \overline{-L,L}} \subset \RM,
\end{equation} 
whose points are labeled in their increasing order. We always center the lattice such that $p_0$ sits at the origin of the real axis. The number of points, {i.e.} the cardinal of $\Ll$, will be denoted by $|\Ll|$ and this number will be assumed infinite in our theoretical analysis but, of course, it will be finite in the numerical simulations. The central assumption of our work is that the points $p_n$ of the lattice are not rendered periodically. Instead, they are generated with the algorithm
\begin{equation}\label{Eq:PattAlg}
p_n = n + r\, \big (\sin[2 \pi (n\theta + \varphi)] -\sin(2 \pi \varphi)\big ), \quad n \in \ZM,
\end{equation} 
where the parameters belong to the circle, $\theta$, $\varphi \in \RM/\ZM$. Our main focus is on the cases when $\theta$ is fixed at irrational values and the lattice is truly aperiodic. The parameter $\varphi$ plays the role of the phason for this pattern. The amplitude $r$ will be fixed at $r=0.45$, such that the points remain ordered with respect to $n$, $p_n < p_{n+1}$. In fact, the pattern of points should be seen as a locally distorted perfect lattice. To fulfill our previous convention, the expression was carefully tailored such that the point $p_0$ corresponding to $n=0$ sits at the origin for all allowed values of the coefficients. A sample of such pattern is shown in Fig.~\ref{Fig:Pattern}. As one can see, with the value $r=0.45$, the pattern is quite far from being periodic.

\vspace{0.2cm}

As it is the case with any aperiodic point pattern \cite{ForrestMAM2002}, the analytic analysis rests on a certain natural dynamical system, which we now describe.  First, note that, for an infinite chain, there exists a natural action of the group $\ZM$ on $\Ll$ on the point patterns given by
\begin{equation}
\ZM \ni a \mapsto \tau_a \Ll = \{ p'_n\}_{n \in \ZM}, \quad p'_n = p_{n+a}-p_a.
\end{equation}
Translated in words, this action shifts the lattice rigidly until the point with the old index $a$ sits at the origin. Note that after the shift, all points are relabeled and the point labeled by $n=0$ sits again at the origin. From \eqref{Eq:PattAlg}, one can see that this action is equivalent to the transformation
\begin{equation}
\varphi \mapsto \varphi + a \theta, \quad a \in \ZM.
\end{equation}
Since the phason $\varphi$ leaves on the circle $\SM = \RM/\ZM$, we can see that the rigid shifts of $\Ll$ translate into rotations by $\theta$ of the phason space $\SM$. Let us point out that all the above conclusions remain the same if the sine function in \eqref{Eq:PattAlg} is replaced with any other continuous function on $\SM$. In fact, as we shall see, our entire analysis rests on the dynamical system $(\SM,\tau)$ indentified as the phason space, hence it applies to an extremely large family of patterns.

\vspace{0.2cm}

Another important observation is that, if $\theta$ is irrational, then the orbit $\{(\varphi + a \theta) \, {\rm mod}\, 1, \, a \in \ZM\}$ of any $\varphi$ under the action of $\ZM$ fills the circle densely. In other words, the dynamical system $(\SM,\tau)$ is topologically ergodic or minimal. This observation plays an important role for the following reason. Let $f,g : \SM \rightarrow \CM$ be any two continuous complex functions over the circle and consider the Hilbert space $\ell^2(\ZM)$ of square-summable sequences over $\ZM$. We associate to $f$ the following diagonal operator over $\ell^2(\ZM)$:
\begin{equation}\label{Eq:PhiRep}
W_f \mapsto \pi_\varphi(f) = \sum_{n \in \ZM} f(\varphi + n \theta) \, |n\rangle \langle n |,
\end{equation}
where $\{|n\rangle \}_{n\in \ZM}$ is the canonical basis of $\ell^2(\ZM)$. If we repeat the same construction for $g$ as well as for the point-wise product $f g$, then it is straightforward to verify that
\begin{equation}
W_f \, W_g = W_{fg} \leftrightarrow \pi_\varphi (f) \pi_\varphi(g) = \pi_\varphi(f g).
\end{equation}
In other words, $\pi_\varphi$ is a representation of the algebra $C(\SM)$ of continuous functions over $\CM$. If $\theta$ is irrational, and only in this case, the representation is faithful and, as such, the algebra generated by the operators \eqref{Eq:PhiRep} is isomorphic to the algebra $C(\SM)$. Furthermore, up to isomorphisms, the $\pi_\varphi$ representations are independent of the parameter $\varphi$. As we shall see, this has important consequences for the spectra of physical operators, particularly, for their independence on $\varphi$. Let us stress that none of these would be true if $\theta$ were rational.

\subsection{The spin system defined}

We will consider a quantum system where one $\frac{1}{2}$-spin sits on each point of the pattern $\Ll$. Hence, the Hilbert space for the spin systems over $\Ll$ is
\begin{equation}
\Hh_S = \bigotimes_{x \in \Ll_L} \CM^2 \simeq M_{D \times D}(\CM^2), \quad D=2^{2L+1}.
\end{equation} 
For aperiodic patterns like the one shown in Fig.~\ref{Fig:Pattern}, it is natural to assume that the spin-spin interaction depends on the separation distance $d_n = p_{n+1}-p_n$ between the spins. As such, we consider spin Hamiltonians of the form
\begin{equation}\label{Eq:GenHam}
H_S = \sum_{n = -L}^L \left [J(d_n) \big (S^{x}_n S^x_{n+1} + S^{y}_n S^y_{n+1} \big ) + J_z(d_n) S^{z}_n S^z_{n+1} \right ] ,
\end{equation}
where
\begin{equation}\label{Eq:SpinGen}
S^\alpha_n= I \otimes \ldots \otimes I \otimes \tfrac{1}{2}\sigma^\alpha \otimes I \ldots \otimes I, \quad \alpha = x,y,z,
\end{equation}
with the Pauli matrix $\sigma^\alpha$ sitting at position $n \in \{ - L,\ldots,L\}$. Both, closed and open boundary conditions will be considered and, to make a clear distinction, we will use $\widehat H$ to indicate the open boundary conditions.

\vspace{0.2cm}

Our methods can handle any functional form of $J$ that is asymptotically decaying and continuous. However, for concreteness, we made the choice 
\begin{equation}
J(s) = e^{-|s|}, \quad s \in \RM,
\end{equation}
which is uniformly used in our numerical simulations. Let us point out that, if we introduce the function
\begin{equation}\label{Eq:F}
f(\varphi) = J\Big (\big |1+ r(\sin[2\pi(\varphi + \theta)]-\sin[2\pi \varphi])\big |\Big),
\end{equation}
defined over the unit circle, then
\begin{equation}
J(d_n) = f(\varphi + n\theta).
\end{equation}
This supplies one connection with the discussion at the end of previous sub-section. As for the Ising interaction strength $J_z$, we chose them to be independent of $d_n$ in order to clearly separate aperiodic from correlation effects.

\vspace{0.2cm}

To fix the notation, let us recall the following relevant operators
\begin{equation}
S^\pm_n = S^x_n \pm \imath S^y_n, \quad n=-L,\ldots,L,
\end{equation}
and the operator of $z$-component of magnetization
\begin{equation}
M = \sum_{n=-L}^L \big ( S_n^z+\tfrac{1}{2} \big ),
\end{equation}
which commutes with the Hamiltonian \eqref{Eq:GenHam}. Also, we recall that the algebra of spin operators accepts the unique trace
\begin{equation}
\Tt_S\big ( \bigotimes_{x \in \Ll} A_x \big )= \prod_{x \in \Ll} \tfrac{1}{2} {\rm Tr}(A_x),
\end{equation}
where ${\rm Tr}$ is the ordinary trace on the $2 \times 2$ matrices. This trace is normalized, $\Tt_S(1\otimes \ldots \otimes 1) =1$.

\vspace{0.2cm}

We now focus on the covariant property of the Hamiltonians and its consequences. With parameter $\theta$ fixed, the Hamiltonians have a functional dependency on $\varphi$, which is now written out explicitly as $H(\varphi)$. Then, if $T_a$ with $a\in \ZM$ represent the usual translations of the spins, we have the obvious covariance relation
\begin{equation}
T_a H(\varphi) T_a^\dagger = H(\varphi + a\theta).
\end{equation}
Since the spectra are invariant under unitary transformations, it follows that ${\rm Spec}(H(\varphi+a\theta))$ are all the same for all $a\in \ZM$. Recall that, if $\theta$ takes an irrational value, then the orbit $\{\varphi+a\theta, \ a\in \ZM\}$ fills the phason space $\SM$ densely. These facts together with the continuity of the spectrum w.r.t. the phason for the aperiodic patterns \cite{FootnoteC} tells us that the spectrum of $H(\varphi)$ is completely independent of $\varphi$. We warn the reader that this remarkable conclusion dose not hold if $\theta$ is rational, a fact that can be easily verified numerically. Furthermore, the covariance relation breaks down in the presence of a boundary, hence the boundary spectrum becomes dispersive w.r.t. the phason.

\vspace{0.2cm}

The characteristics described above are ideal for topological Thouless pumping. Let us recall that, in general, it is quite difficult to design Hamiltonians that depend on a parameter in such a way that at least one bulk gap does not close as the parameter is cycled. The above discussion tells us that all the bulk spectral gaps of $H(\varphi)$ remain unchanged, hence open, when the phason is varied. This is a remarkable property unmatched by any other design method and this is why the quasi-periodic systems are so valuable for practical applications.

\subsection{Connections with fermionic models}

We will use the Jordan-Wigner mapping \cite[Sec.~5.1]{ColemanBook} to make connections with fermionic aperiodic physical systems already studied in the literature. We want to make clear from the beginning, however, that the fermionic models will only be used in the theoretical analysis. The numerical analysis is always performed with the spin Hamiltonian \eqref{Eq:GenHam}.

\vspace{0.2cm}

\vspace{0.2cm}

For the reader's convenience, we recall that the Jordan-Wigner mapping is supplied by the operators
\begin{equation}\label{Eq:StoA}
a_n = S_n^- \prod_{j=-L}^{n-1} 2S_j^z, \quad a_n^\ast = S_n^+ \prod_{j=-L}^{n-1} 2S_n^z,
\end{equation}
with $n$ running from $-L$ to $L$. These operators satisfy the canonical anti-commutation relations
\begin{equation}\label{Eq:Rel2}
a_m a_n + a_n a_m=0, \quad a_m^\ast a_n + a_n a_m^\ast = \delta_{m,n}.
\end{equation} 
Together with the inverse formulas
\begin{equation}\label{Eq:AtoS}
S_n^{-} = a_n \prod_{j=-L}^{n-1}(2 a_j^\ast a_j -1), \ \  S_n^{+} = a_n^\ast \prod_{j=-L}^{n-1}(2 a_j^\ast a_j -1),
\end{equation}
these relations establish an isomorphism between the algebra of spin operators and the algebra of fermionic creation and annihilation operators over the lattice $\Ll$. In particular, the magnetization operator is mapped into the particle number
\begin{equation}
M \rightarrow N=\sum_{n=-L}^L a_n^\ast a_n,
\end{equation}
and the model Hamiltonian \eqref{Eq:GenHam} into \cite[p.~74]{ColemanBook}
\begin{align}\label{Eq:FHam}
H_F = \sum_{n = -L}^L & \big [\tfrac{1}{2} J(d_n) (a^\ast_n a_{n+1} + a^\ast_{n+1} a_{n} ) \\ \nonumber
& +J_z \sum_{n =-L}^L (a_n^\ast a_n -\tfrac{1}{2})(a_{n+1}^\ast a_{n+1} -\tfrac{1}{2}) \big ],
\end{align}

The algebra of fermion operators also accepts a unique normalized trace, to be denoted by $\Tt_F$. The Jordan-Wigner transformation preserves the unique traces of the two algebras.

\section{The Non-Commutative $d$-Torus}
\label{Sec:NCT}

There is a natural link between the point pattern in \eqref{Eq:PattAlg} and the algebra called non-commutative torus \cite{ProdanJGP2019}. In dimensions $d=2$ and $d=3$, this algebra coincides with the algebra of magnetic translations and this fact has been used to fabricate patterned meta-materials that mimic the physics of the integer quantum Hall effect \cite{ApigoPRM2018}.

\vspace{0.2cm}

As we shall see, when we restrict the spin model to the invariant subspace of the magnetization operator, the reduced algebra of physical observables can be computed explicitly and it coincides with the non-commutative $d$-torus, where the dimension $d$ is determined by the value of the magnetization. In this section, we review the basic facts about this algebra.

\subsection{The non-commutative $d$-torus defined}

Let $\Theta=\{\theta_{ij}\}_{i,j=\overline{1,d}}$ be a $d\times d$ antisymmetric matrix with entries from $\RM/\ZM$. The non-commutative $d$-torus associated to $\Theta$ is the universal $C^\ast$-algebra
\begin{equation}
\Aa_\Theta = C^\ast(u_1,\ldots,u_d),
\end{equation}
generated by $d$-unitary elements satisfying the relations
\begin{equation} 
u_i u_j=e^{\imath 2 \pi \theta_{ij}}u_ju_i, \quad i,j = 1, \ldots,d.
\end{equation}
A generic element of the algebra can be presented in the following form
\begin{equation}
a = \sum_{\bm q \in \ZM^d} a_{\bm q}\, u_{\bm q}, \quad u_{\bm q} = u_1^{q_1} \ldots u_d^{q_d}, \quad a_{\bm q} \in \CM,
\end{equation}
but other conventions are possible. When all entries of $\Theta$ are irrational and rationally independent, the non-commutative torus accepts a unique trace
\begin{equation}\label{Eq.NCTrace}
\Tt \Big ( \sum_{\bm q \in \ZM^d} a_{\bm q}\, u_{\bm q} \Big ) = a_{\bm 0}.
\end{equation}
The monomials $u_{\bm n}$ are orthonormal with respect to the scalar product induced by the trace
\begin{equation}
\langle u_{\bm n},u_{\bm n'}\rangle := \Tt\big (u_{\bm n}^\ast u_{\bm n'}\big ) = \delta_{\bm n,\bm n'}, \quad \bm n,\bm n' \in \ZM^d,
\end{equation}
and $\big (\Aa_\Theta, + , \langle , \rangle \big )$ becomes a Hilbert space on which the elements of the algebra act as
\begin{equation}
\pi(a)|a'\rangle = |a a' \rangle, \quad a,a' \in \Aa_\Theta.
\end{equation}
If we use the shorthand $|\bm n \rangle$ for $|u_{\bm n}\rangle$, then it is straightforward to see that this Hilbert space is just $\ell^2(\ZM^d)$, the space of square summable sequences labeled by $\ZM^d$. Furthermore,
\begin{equation}\label{Eq:MagTr}
\pi(u_{\bm q})|\bm n \rangle = |u_{\bm q} u_{\bm n} \rangle = e^{2 \pi \imath \langle \bm q|\Theta_+|\bm n\rangle}|u_{\bm q + \bm n}\rangle,
\end{equation}
where $\Theta_+$ is the upper diagonal part of $\Theta$. Eq.~\eqref{Eq:MagTr} is just the magnetic translation by $\bm q$ in ordinary tight-binding solid state models, written in the Landau gauge. In this representation, the entries $\theta_{ij}$ of $\Theta$ correspond to the flux of the magnetic field through the facet $\{i,j\}$ of the primitive cell, expressed in half the quantum of flux unit $h/2e$. 

\vspace{0.2cm}

The above representation, which is just the standard Gelfand-Naimark-Segal representation\cite{DavidsonBook} of $\Aa_\Theta$ induced by the trace $\Tt$, connects this algebra with the algebra of magnetic translations. In this work, however, we will encounter different representations of $\Aa_\Theta$. Nevertheless, being the same algebra, the spectra of the Hamiltonians, at least for $d=2$, resemble quite closely the Hofstadter butterfly \cite{Hoftadter1976} seen in the spectrum of 2-dimensional electrons in magnetic fields.

\subsection{Elements of $K$-theory}
\label{Ch:KTh}

In this work, we use the complex K-theory of operator algebras \cite{BlackadarBook1}, which is a natural extension of the K-theory of vector bundles \cite{ParkBook}. This theory supplies all independent topological invariants that can be associated to projections and unitary elements of an algebra.

\vspace{0.2cm}

The complex $K$-theory of  the algebra $\Aa_\Theta$ contains two $K$-groups, which can be described as follows. The first one is the $K_0(\Aa_\Theta)$ group, which classifies the projections
\begin{equation}
p \in \Mm_\infty \otimes \Aa_\Theta, \quad p^2 = p^\ast=p,
\end{equation}
with respect to the von~Neumann equivalence relation
\begin{equation}\label{Eq-EquivRelation}
p \sim p' \quad \mbox{iff}  \quad p=vv' \ \  {\rm and} \ \ p' = v'v, 
\end{equation}
for some partial isometries $v$ and $v'$ from $\Mm_\infty \otimes \Aa_\Theta$. Above, $\Mm_N$ is the algebra of $N \times N$ matrices with complex entries and $M_\infty$ is the direct limit of these algebras. For any projection $p$ from $\Mm_\infty \otimes \Aa_\Theta$, there exists $N \in \NM$ such that $p \in \Mm_N \otimes \Aa_\Theta$, hence we do not really need to work with infinite matrices. However, $\Mm_N$ can be canonically embedded into $\Mm_\infty$ and this convenient, because it enables $N$ to take flexible values.

\vspace{0.2cm}

We need to answer two questions: 1) How does the equivalence relation \eqref{Eq-EquivRelation} supply topological information? 2) Why do we need the tensoring by $\Mm_\infty$? Both questions find their answers in the following remark. There are two additional equivalence relations for projections \cite[p.~18]{ParkBook}:
\begin{itemize}
\item  Similarity equivalence:
\begin{equation}
p \sim_u p' \quad {\rm iff} \quad p'= u p u^\ast
\end{equation}
for some unitary element $u$ from $\Mm_\infty \otimes \Aa_\Theta$;
\item Homotopy equivalence:
\begin{equation}
p \sim_h p' \quad  {\rm iff} \quad \bm p(0)=p \ \  {\rm and} \ \  \bm p(1) = p'
\end{equation}
for some continuous function $\bm p : [0,1] \rightarrow \Mm_\infty \otimes \Aa_\Theta$, which always returns a projection. 
\end{itemize}
The homotopy equivalence is certainly the topological equivalence as understood by condensed matter physicists. Now, in general, the three equivalence relations are different, but 
tensoring $\Aa_\Theta$ by $\Mm_\infty$ makes them entirely equivalent. For topological classification, $\sim_h$ is the most interesting relation, but, as we shall see, the relation $\sim$ is essential for understanding the spectral properties of Hamiltonians. 

\vspace{0.2cm}

The equivalence class of a projection $p$ will be denoted by $[p]_0$, hence, $[p]_0$ is the set
\begin{equation}
[p]_0= \big \{p' \in \Mm_\infty \otimes \Aa_\Theta \, ,  \ p' \sim p \big \}.
\end{equation}
If $p \in \Mm_N \otimes \Aa_\Theta$ and $p' \in \Mm_M \otimes \Aa_\Theta$ are two projections, then $\begin{pmatrix} p & 0 \\ 0 & p' \end{pmatrix}$ is a projection from $\Mm_{N+M} \otimes \Aa_\Theta$ and one can define the addition
\begin{equation}
[p]_0 \oplus [p']_0 = \left [ \begin{matrix} p & 0 \\ 0 & p' \end{matrix} \right ]_0,
\end{equation}
which provides a semigroup structure on the set of equivalence classes. Then $K_0(\Aa_\Theta)$ is its  enveloping group  \cite{BlackadarBook1} and, for the non-commutative $d$-torus,
\begin{equation}\label{Eq:K0}
K_0(\Aa_\Theta) = \ZM^{2^{d-1}},
\end{equation}
regardless of $\Theta$. As such, there are $2^{d-1}$ generators $[e_J]_0$, which can be uniquely labeled by the subsets of indices $J \subseteq \{1,\ldots,d\}$ of even cardinality \cite{ProdanSpringer2016}. Throughout, the cardinality of a set will be indicated by $|\cdot |$. Eq.~\eqref{Eq:K0} assures us that, for any projection $p$ from $\Mm_\infty \otimes \Aa_\Theta$, one has
\begin{equation}\label{Eq-GenExpansion}
[p]_0  = \sum_{J \subseteq \{1,\ldots,d\}}^{|J|={\rm even}} n_J \, [e_J]_0, 
\end{equation} 
where the coefficients $n_J$ are integer numbers that do not change as long as $p$ is deformed inside its $K_0$-class. Specifically, two homotopically equivalent projections will display the same coefficients, hence $\{n_J\}_{|J|={\rm even}}$ represent the {\it complete} set of topological invariants associated to the projection $p$. Furthermore, two projections that display the same set of coefficients are necessarily in the same $K_0$-class. Let us point out that the coefficient $n_J$ corresponding to $J=\{1,2,\ldots,d\}$ is called the top coefficient and is equal to the strong Chern number associated to the projection $p$ \cite[Sec.~5.7]{ProdanSpringer2016}.

\vspace{0.2cm}

The second group of the complex $K$-theory is $K_1(\Aa_\Theta)$, which classifies the unitary elements
\begin{equation}
u \in \Mm_\infty \otimes \Aa_\Theta, \quad u u^\ast = u^\ast u=1,
\end{equation}
with respect to the homotopy equivalence relation. The class of $u\in \Mm_\infty \otimes \Aa_\Theta$ will be denoted by $[u]_1$. For the non-commutative $d$-torus,
\begin{equation}
K_1(\Aa_\Theta) = \ZM^{2^{d-1}},
\end{equation}
regardless of $\Theta$. Again, there are $2^{d-1}$ generators $[u_J]_1$, which can be uniquely labeled by the subsets of indices $J \subseteq \{1,\ldots,d\}$ of odd cardinality \cite{ProdanSpringer2016}. This assures us that, for any unitary $u$ from $\Mm_\infty \otimes \Aa_\Theta$, one has
\begin{equation}
[u]_1  =\sum_{J \subseteq \{1,\ldots,d\}}^{|J|={\rm odd}} n_J \, [u_J]_1,
\end{equation} 
and the coefficients $n_J$ are again integer numbers that do not change as long as $u$ is deformed inside its class. Specifically, two homotopic unitaries will display the same coefficients, hence $\{n_J\}_{|J|={\rm odd}}$ represent the complete set of topological invariants associated to $u$.  

\subsection{Relation to Chern numbers}

We will use the uniform notation from \cite{ProdanSpringer2016} for the weak and the strong Chern numbers of a gap projection, namely, ${\rm Ch}_J(P_G)$, where $J \subset \{1,\ldots,d\}$ is a subset of directions. The values of the Chern numbers on the $K_0$-generators can be found in \cite{ProdanSpringer2016}[p.~141]:
\begin{equation}\label{Eq:ChernValues}
{\rm Ch}_{J'} [e_J]_0 = \left \{ 
\begin{array}{l}
0 \ {\rm if} \ J'\nsubseteq J  , \\
1 \ {\rm if} \ J' = J , \\
{\rm Pf}(\Phi_{J\setminus J'}) \ {\rm if} \ J' \subset J,
\end{array}
\right .  \quad J, J' \subset \{1,\ldots,d\}.
\end{equation}
Since the Chern numbers are also linear maps, their values on the gap projection $[P_G]_0 = \sum_{J} n_J \, [e_J]_0$ can be straightforwardly computed from \eqref{Eq:ChernValues}:
\begin{equation}\label{Eq:ChernVal}
{\rm Ch}_{J'}[P_G]_0 = n_{J'} + \sum_{J' \subsetneq J} n_J \, {\rm Pf}(\Phi_{J\setminus J'}).
\end{equation} 
Let us point out that the top Chern number corresponding to $J'=\{1,\ldots,d\}$ is always an integer, but the lower Chern numbers may not be.

\subsection{Spectral gap labeling}\label{Sec-GapLabeling}

Let $h \in \Mm_\infty \otimes \Aa_\Theta$ be a Hermitian element and $G$ a gap in its spectrum. Depending on the context, the symbol $G$ will stand for the energy interval or for the center of this interval. Let $\chi(s)$ be the step function which drops from 1 to 0 at $s=0$. Using functional calculus, we can define the gap projection $p_G \equiv \chi(h - G)$. Being a projection from $\Mm_\infty \otimes \Aa_\Theta$, it defines an equivalence class in $K_0(\Aa_\Theta)$ and, per previous discussion, we have the decomposition
\begin{equation}
[p_G]_0 = \sum_{J \subseteq \{1,\ldots,d\}}^{|J|={\rm even}} n_J(G) \,  [e_J]_0.
\end{equation}
If $G'$ is another spectral gap of $h$, then $p_G$ and $p_{G'}$ cannot be homotopically connected, hence they belong to different $K_0$-classes and, as such, the two projectors will display different sets of integer coefficients $\{n_J\}$. The conclusion is that the spectral gaps are uniquely labeled by the $K_0$-group itself.  This principle was discovered by Jean Bellisssard in his pioneering applications of $K$-theory to solid state physics \cite{Bellissard1986,Bellissard1995}.

\vspace{0.2cm}

\begin{figure}[t!]
\includegraphics[width=\linewidth]{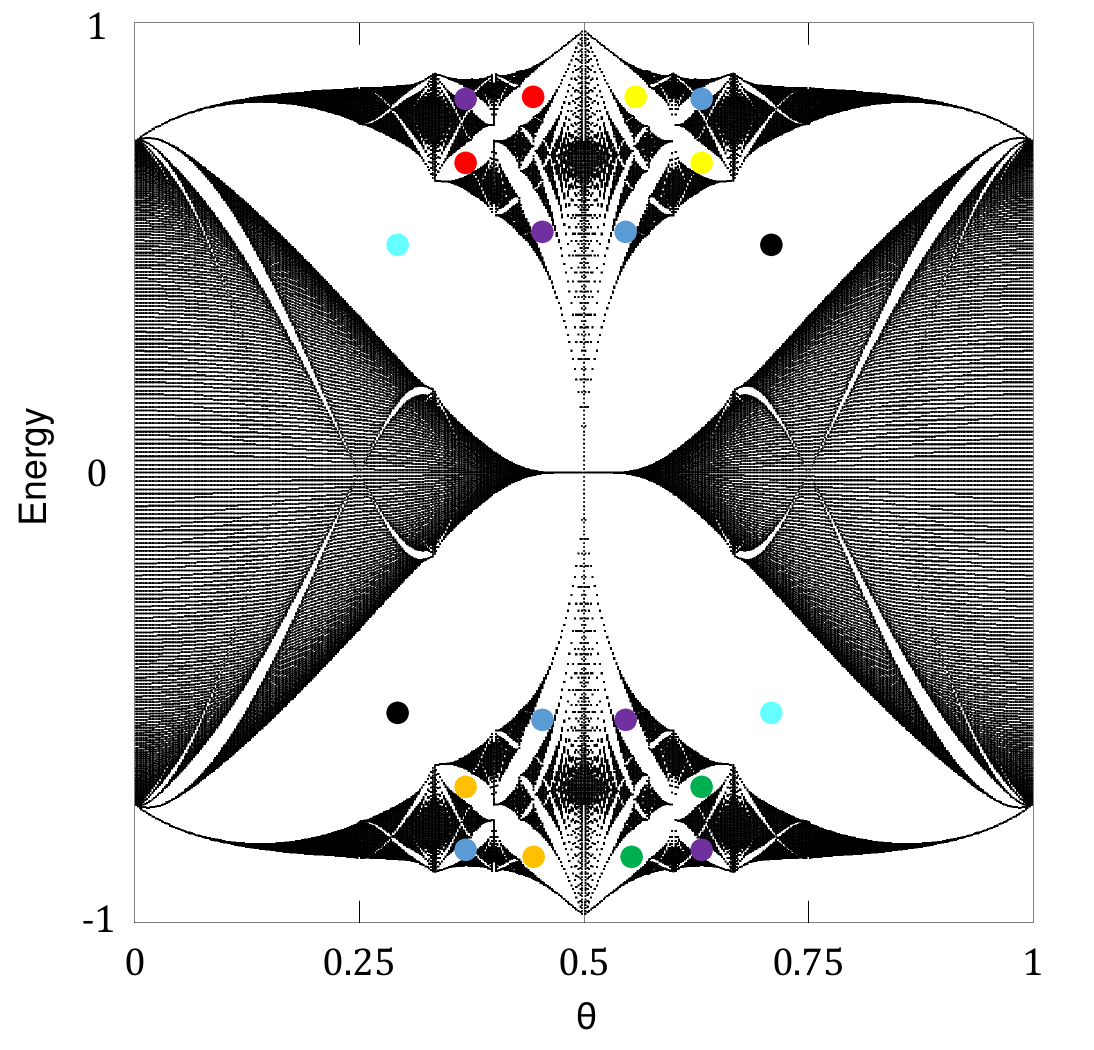}
\caption{\small Energy spectrum of Hamiltonian~\eqref{Eq:GenHam} inside the $M=1$ sector as function of $\theta$. The simulations were performed with for $J_z=0$, $|\Ll|=501$ and the range of parameter $\theta$ has been sampled at rational values $\theta_n = \frac{n}{|\Ll|}$ to accommodate closed boundary conditions. Eight gaps are identified and color coded for future references.}
\label{Fig:M1JZ0Spec}
\end{figure}

To add more clarity to the above statement, let us recall that, when one enumerates the elements of a set, one actually assigns labels using elements from the group $\ZM$. For the set of spectral gaps in the Hofstadter butterfly (see Fig.~\ref{Fig:M1JZ0Spec}), if one tries to count, say, starting from the bottom of the spectrum, one will soon realize that it is impossible, because between any two spectral gaps there are an infinite number of additional spectral gaps. What we were asserting in the previous paragraph was that one needs to count the spectral gaps not by $\ZM$ but by the $K_0$-group. Furthermore, when looking at the Hofstadter butterfly, what really jumps to ones eyes is the structure and the pattern of the spectral gaps and not the quantitative details. It is fair to say that, when looking at the spectrum of a class of Hamiltonians, more precisely at the structure of the gaps, we literally see a representation of the $K_0$-group of the algebra which contains those Hamiltonians. By diagonalizing more and more Hamiltonians, we can look at the $K_0$-group from ``different angles'' and ultimately we can identify it entirely.

\vspace{0.2cm}

The gap labels can be detected numerically by a variety of methods. For example, there are precise relations between the strong and weak Chern numbers and the gap labels \cite[Sec.~5.7]{ProdanSpringer2016}, which generalize the well known Streda formula \cite{StredaJPC1982}. However, one of the most effective tools is supplied by the integrated density of states (IDS) of the Hamiltonians over $\ell^2(\ZM^d)$
\begin{equation}\label{Eq:IDS1}
{\rm IDS}(E) = \lim_{V \rightarrow \RM^d}\frac{\big | {\rm Spec}\big (\Pi_V \, H \, \Pi_V \big ) \cap (-\infty,E] \big |}{\big |V \cap \ZM^d \big |},
\end{equation}
where $|\cdot|$ denotes the cardinal of a set and $\Pi_V$ represents the projection onto the sites $\bm q \in V$. Translated in words, ${\rm IDS}(E)$ is the number of eigenvalues below $E$ of any finite-volume representation of the Hamiltonian, divided by the number of sites inside that volume, for large enough volumes. Note that $\Pi_V H\Pi_V$ is just the bulk Hamiltonian restricted on $V$ via  Dirichlet boundary condition, but any other boundary condition will do. 

\vspace{0.2cm}

Definition \eqref{Eq:IDS1} is very convenient for numerical evaluations. However, the topological information encoded by IDS is revealed by another expression. Indeed, when $E$ belongs to a spectral gap $G$, then IDS can be equivalently computed as
\begin{equation}\label{Eq:IDS2}
{\rm IDS}(G) = {\rm Tr}_V(P_G),
\end{equation}
where $P_G$ is the gap projector and ${\rm Tr}_V$ is the trace per volume over the Hilbert space $\ell^2(\ZM^d)$,
\begin{equation}\label{Eq:TrV}
{\rm Tr}_V(P_G) = \lim_{V \rightarrow \RM^d}\frac{ {\rm Tr}(P_G)}{\big |V \cap \ZM^d \big |}.
\end{equation}
 Now, if $H$ is a physical representation of an element $h \in \Aa_\Theta$, $H=\pi(h)$, then
\begin{equation}
{\rm Tr}_V(P_G) = \Tt(p_G), \quad P_G = \pi(p_G),
\end{equation}
where $\Tt$ is the trace introduced in Eq.~\eqref{Eq.NCTrace}. The above relation is well known and is a direct consequence of Birkhoff ergodic theorem \cite{Birkhoff1931}. Consider now a topological deformation $p'_G$ of $p_G$, which can be induced by a deformation of the Hamiltonian itself. Definitely, these two projections belong to the same $K_0$ classes, hence they are connected by two partial isometries $p_G = vv'$ and $p'_G=v'v$. Since any trace is invariant to cyclic permutations of the entries, one finds that $\Tt(p_G) = \Tt(p'_G)$. Hence, $\Tt$ is constant over the $K_0$ classes. Furthermore, being a linear map,
\begin{equation}
\Tt\big ([p_G]_0\big ) = \sum_{J \subseteq \{1,\ldots,d\}}^{|J|={\rm even}} n_J(G) \, \Tt\big ([e_J]_0\big ).
\end{equation}

\vspace{0.2cm}

\begin{figure}[t!]
\includegraphics[width=\linewidth]{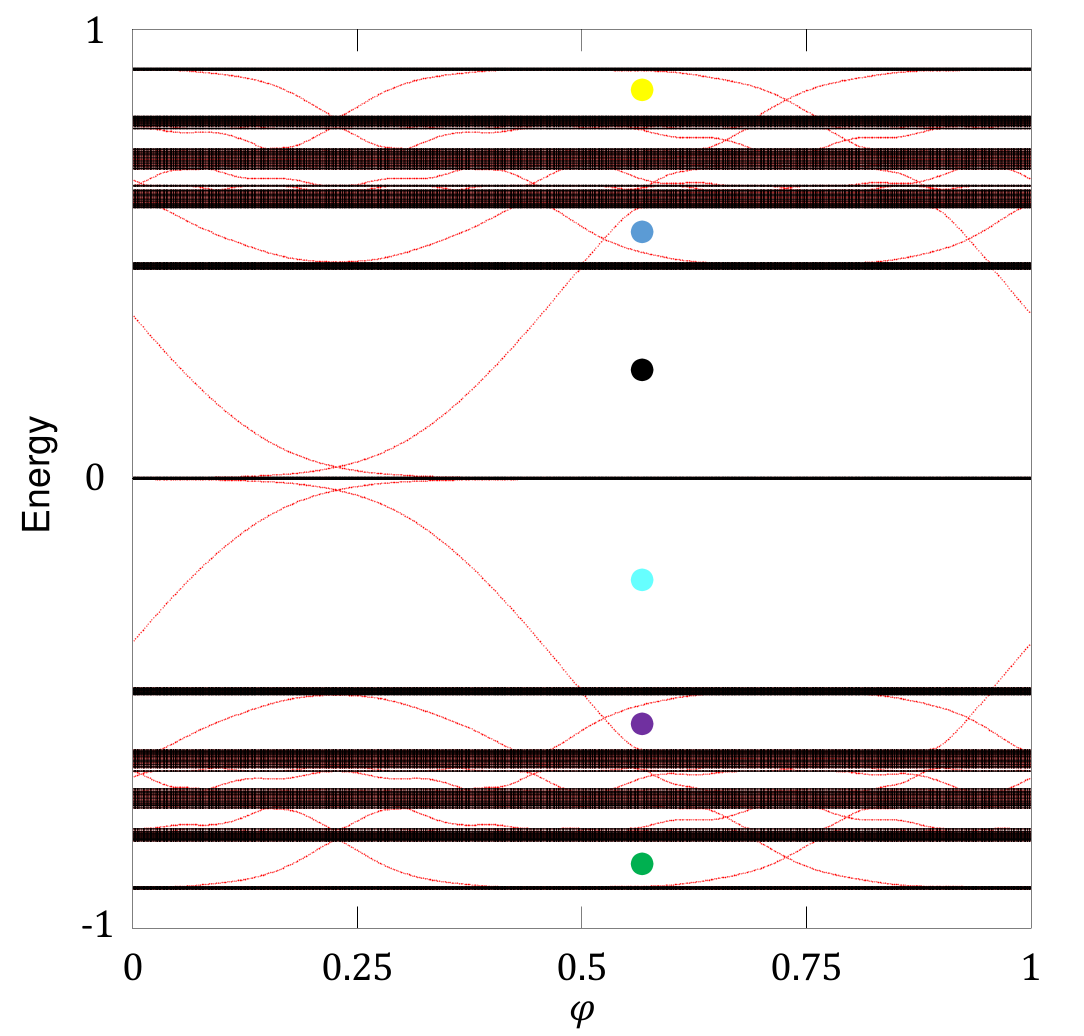}
\caption{\small Energy spectrum of Hamiltonian \eqref{Eq:GenHam} with open boundary condition inside the $M=1$ sector, plotted as function of parameter $\varphi$ from \eqref{Eq:PattAlg}. The topological boundary spectrum is highlighted in red and the spectrum with closed boundary conditions is overlaid in black color. The computation was performed with $J_z=0$, $\theta=\frac{1+\sqrt{3}}{5}$ and $|\Ll|=473$. The colored dots mark the same gaps as in Fig.~\ref{Fig:M1JZ0Spec}.}
\label{Fig:M1JZ0Phi}
\end{figure} 

The values of the trace on the generators $e_J$ were computed in \cite{Elliott1984} (see also \cite[Sec.~5.7]{ProdanSpringer2016}):
\begin{equation}
\Tt\big ( [e_J]_0\big ) = {\rm Pfaff}(\Theta_J),
\end{equation}
where on the right we have the Pfaffian of the matrix obtained from $\Theta$ by restricting to the indices contained in the subset $J$. The conclusion is that we can predict the range of the IDS when evaluated inside the spectral gaps
\begin{equation}\label{Eq:MainIDS}
{\rm IDS}(G) = \Tt\big ([p_G]_0\big ) = \sum_{J \subseteq \{1,\ldots,d\}}^{|J|={\rm even}} n_J(G) \, {\rm Pfaff}(\Theta_J).
\end{equation}
When the entries of $\Theta$ are rationally independent, all coefficients $n_J(G)$ can be detected from the values of the IDS. In fact, in such situations, the IDS supplies a group isomorphism between $K_0(\Aa_\Theta)$ and a dense but nevertheless countable subgroup of $\RM$.

\section{Topological Gaps: The Non-Correlated Case}
\label{Sec:TopoGaps1}

In this section, we set $J_z =0$ and investigate the magnetization sectors separately for up to $M=3$. As we shall see, for all cases, the energy spectrum of the spin Hamiltonian \eqref{Eq:GenHam} displays fractality and one of the goals is to label the spectral gaps of the fractal butterfly by appropriate K-groups. Another goal is to demonstrate the emergence of topological edge modes when the parameter $\varphi$ in Eq.~\eqref{Eq:PattAlg} is varied.

\begin{figure*}[t!]
	\includegraphics[width=\linewidth]{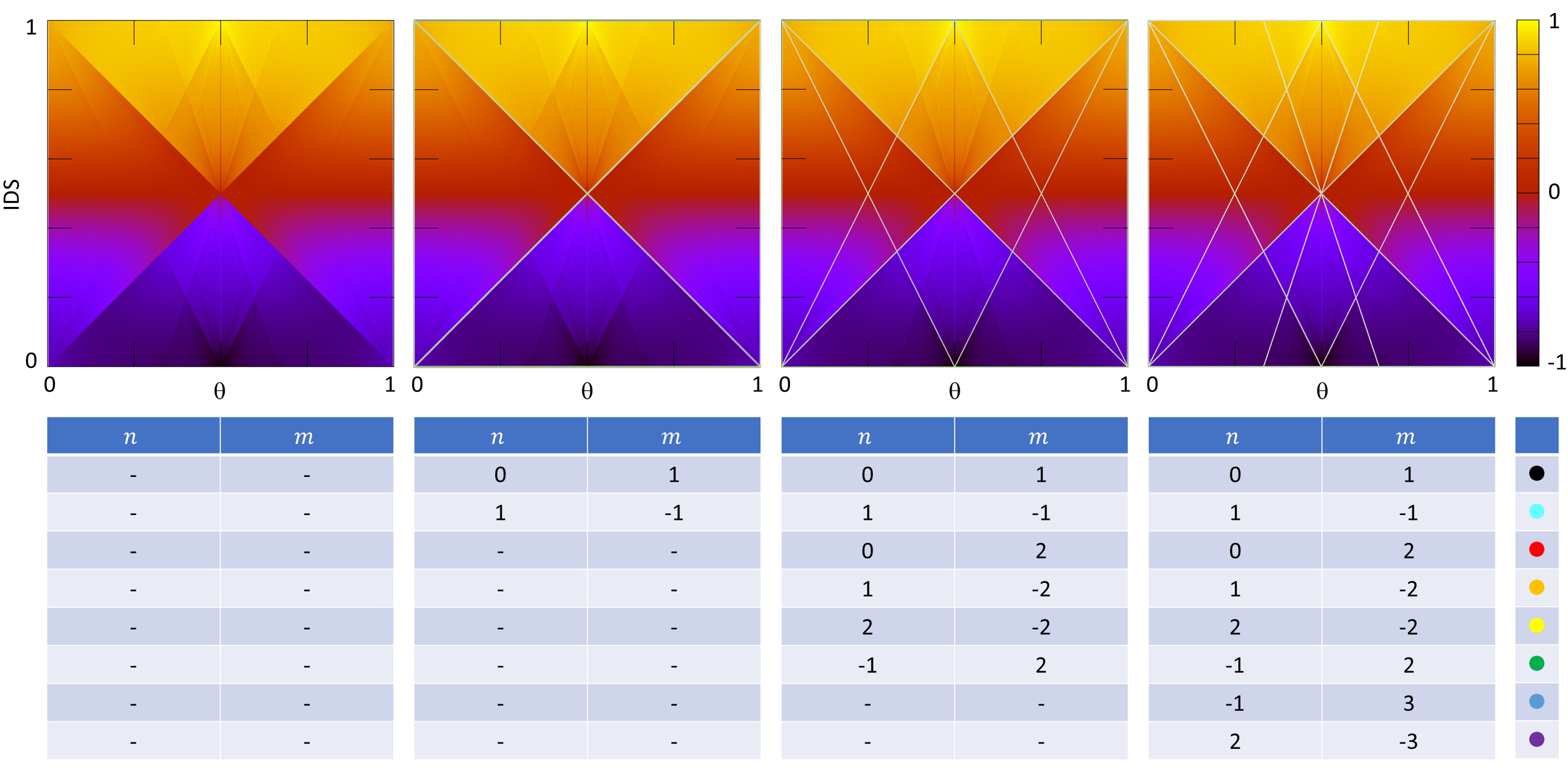}
	\caption{\small Numerical IDS as computed from the spectral butterfly reported in Fig.~\ref{Fig:M1JZ0Spec} for the $M=1$ sector and $J_z=0$. The K-theoretic IDS values from Eq.~\eqref{Eq:IDS1Pred}, shown as light colored lines, are matched with the numerical IDS values inside the gaps marked in Fig.~\ref{Fig:M1JZ0Spec}, identified here by the abrupt changes in the color plot. The matching progresses in the order of the gap sizes. The tables list the gap labels $(n,m)$ from Eq.~\eqref{Eq:IDS1Pred} as well as the corresponding gaps.}
	\label{Fig:GapLabels_M1JZ0}
\end{figure*}

\subsection{The $M =1$ Sector}
\label{Sec:M1Jz0}

In Fig.~\ref{Fig:M1JZ0Spec}, we report the spectrum of the Hamiltonian defined in Eq.~\eqref{Eq:GenHam} as a function of the parameter $\theta$ from Eq.~\eqref{Eq:PattAlg}, computed inside the $M=1$ sector with closed boundary conditions. The resemblance between  that spectrum and the Hofstadter spectrum of the electrons in a magnetic field \cite{Hoftadter1976} is evident, which may appear strange at first sight given the fact that no fine tunning was performed. Figure~
\ref{Fig:M1JZ0Phi} reports the spectrum of the same Hamiltonian computed with open boundary conditions at a fixed $\theta$, but a variable parameter $\varphi$ from Eq.~\eqref{Eq:PattAlg}. The observed chiral bands is an indication that the spectral gaps are topological. This subsection supplies an explanation of both observations based on an explicit computation of the algebra of physical observables and of its K-theory. In the process, we exemplify how the IDS can be used to identify the topological labels of the spectral gaps and how to work out the bulk-boundary correspondence.

\vspace{0.2cm}

The $M=1$ sector of the spin system is very simple. A basis for this sector consists of states with all spins down and only one spin up. We denote such state as $|n\rangle$ if the up-spin is located at site $n$ and the Hilbert space generated by these states by $\Hh_1$. Obviously, $\Hh_1 \simeq \ell^2(\ZM)$. The operators $S^{x}_n S^x_{n+1} + S^{y}_n S^y_{n+1}$ act as simple hopping operators on $\ell^2(\ZM)$ and the spin Hamiltonian reduces to the ordinary tight-binding Hamiltonian
\begin{equation}
H_1 = \tfrac{1}{2}\sum_{n \in \mathbb Z} J(d_n) \, \big (|n\rangle \langle n+1| + |n +1 \rangle \langle n| \big ).
\end{equation}
This expression, which also follows from the Jordan-Wigner transformation, can be re-written in the following form:
\begin{align}\label{Eq:H1}
H_1 = & \tfrac{1}{2} T \sum_{n \in \mathbb Z}  f(\varphi+n\theta) \, |n\rangle \langle n| \\ 
& + \tfrac{1}{2} T^\ast \sum_{n \in \mathbb Z} f(\varphi + (n-1)\theta) \, |n\rangle \langle n|,  
\end{align}
where $T$ is the lattice shift operator $T|n\rangle = |n+1\rangle$ and the function $f$ was supplied in Eq.~\eqref{Eq:F}. Without imposing constraints on the functional dependence $J$, one can see that $H_1$ is generated by the translation operator $T$ and by diagonal operators of the form:
\begin{equation}\label{Eq:DiagOp}
W_g=\sum_{n \in \mathbb Z}  g(\varphi+n\theta) \, |n\rangle \langle n|, \quad g\in C(\SM),
\end{equation} 
already introduced in Eq.~\eqref{Eq:PhiRep}. Furthermore, we have the following commutation relation:
\begin{align}\label{Eq:Comm1}
 & \Big ( \sum_{n \in \mathbb Z}  g(\varphi+n\theta) \, |n\rangle \langle n| \Big ) T \\ \nonumber 
 & \qquad \qquad = T \Big ( \sum_{n \in \mathbb Z}  g(\varphi+(n+1)\theta) \, |n\rangle \langle n| \Big ),
\end{align}
or, more compactly:
\begin{equation}
W_g \, T = T \, W_{g\circ \tau_\theta},
\end{equation}
where $\tau_\theta$ is the rotation of the circle by $\theta$.

\begin{figure}[t!]
\includegraphics[width=\linewidth]{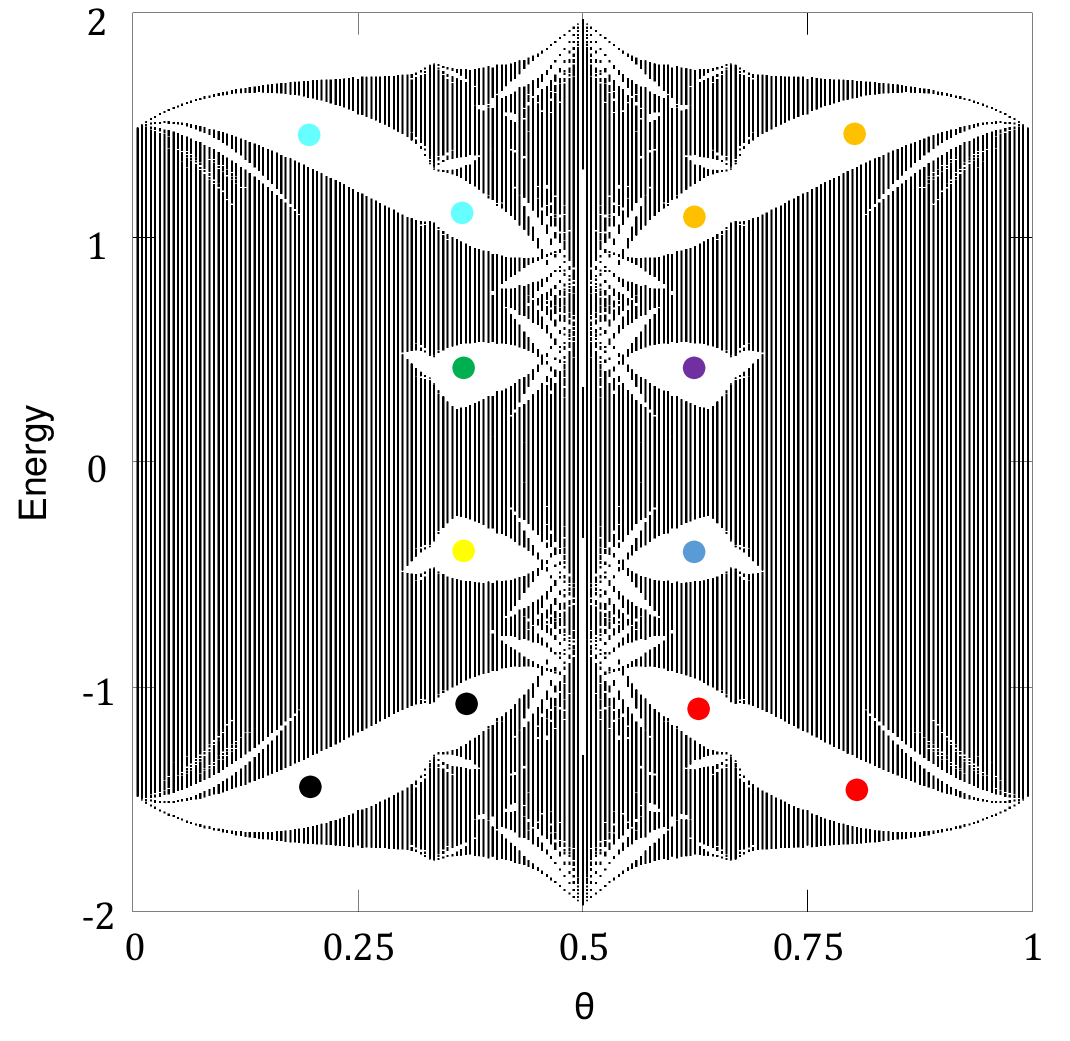}
\caption{\small Energy spectrum of the Hamiltonian~\eqref{Eq:GenHam} as a function of $\theta$ for $M=2$ sector and $J_z=0$. The simulations were performed for a chain with $|\Ll|=201$ and the range of parameter $\theta$ has been sampled at rational values $\theta_n = \frac{n}{|\Ll|}$  to accommodate closed boundary conditions. Eight prominent gaps are identified and color-coded for future references.}
\label{Fig:M2JZ0Spec}
\end{figure}

\vspace{0.2cm}

\begin{figure}[t!]
\includegraphics[width=\linewidth]{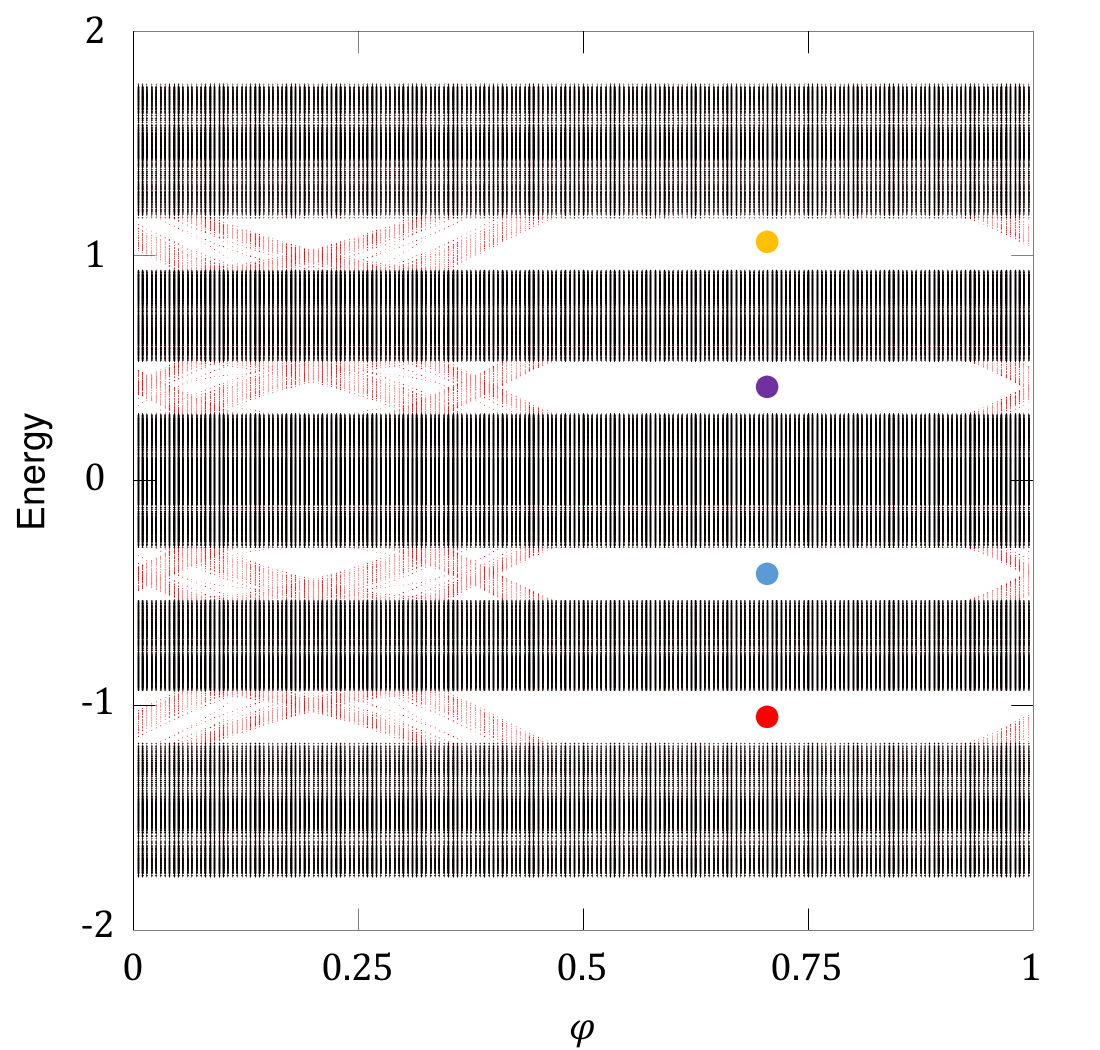}
\caption{\small Energy spectrum of the Hamiltonian \eqref{Eq:GenHam} inside the $M=2$ sector and with open boundary condition, plotted as a function of parameter $\varphi$ from \eqref{Eq:PattAlg}. The topological boundary spectrum is highlighted in red and the spectrum with closed boundary condition is overlaid in black color. The computation was performed with $J_z=0$, $\theta=\frac{1+\sqrt{2}}{4}$ and $|\Ll | = 169$. The colored dots mark the same gaps as in Fig.~\ref{Fig:M2JZ0Spec}.} 
\label{Fig:M2JZ0Phi}
\end{figure}

Per discussion in Sec.~\ref{SSec:AperLattice}, the algebra generated by the operators \eqref{Eq:DiagOp} is isomorphic to the algebra $C(\SM)$ of continuous functions over the circle. Furthermore, any function $f$ over the circle can be Fourier decomposed. As such, the algebra of continuous functions over $\SM$ is generated by a single function:
\begin{equation}
u: \RM/\ZM \rightarrow \CM, \quad u(x) = e^{\imath 2\pi x}.
\end{equation} 
Hence all the diagonal operators from Eq.~\eqref{Eq:DiagOp} can be obtained as linear combinations of powers of a single diagonal operator:
\begin{equation}\label{Eq:U}
U=e^{-\imath 2 \pi \varphi}\sum_{n \in \mathbb Z}  u(\varphi+n\theta) \, |n\rangle \langle n|=\sum_{n \in \mathbb Z}  e^{\imath 2 \pi n \theta} \, |n\rangle \langle n|.
\end{equation}
The conclusion is that, regardless of the functional dependence on $d_n$ of the coupling coefficients, the Hamiltonian $H_1$ is drawn from the algebra $C^\ast(T,U)$ generated by $T$ and $U$ and one can check from Eq.~\eqref{Eq:Comm1} the following commutation relation:
\begin{equation}\label{Eq:Co1}
UT = e^{\imath 2 \pi \theta} TU.
\end{equation}
Hence, the algebra of observables coincides with the non-commutative 2-torus $\Aa_{\Theta_1}$, with a $\theta$-matrix:
\begin{equation}
\Theta_1 = \begin{pmatrix} 0 & \theta \\ - \theta & 0 \end{pmatrix}.
\end{equation}

\vspace{0.2cm}

\begin{figure*}[t!]
	\includegraphics[width=\linewidth]{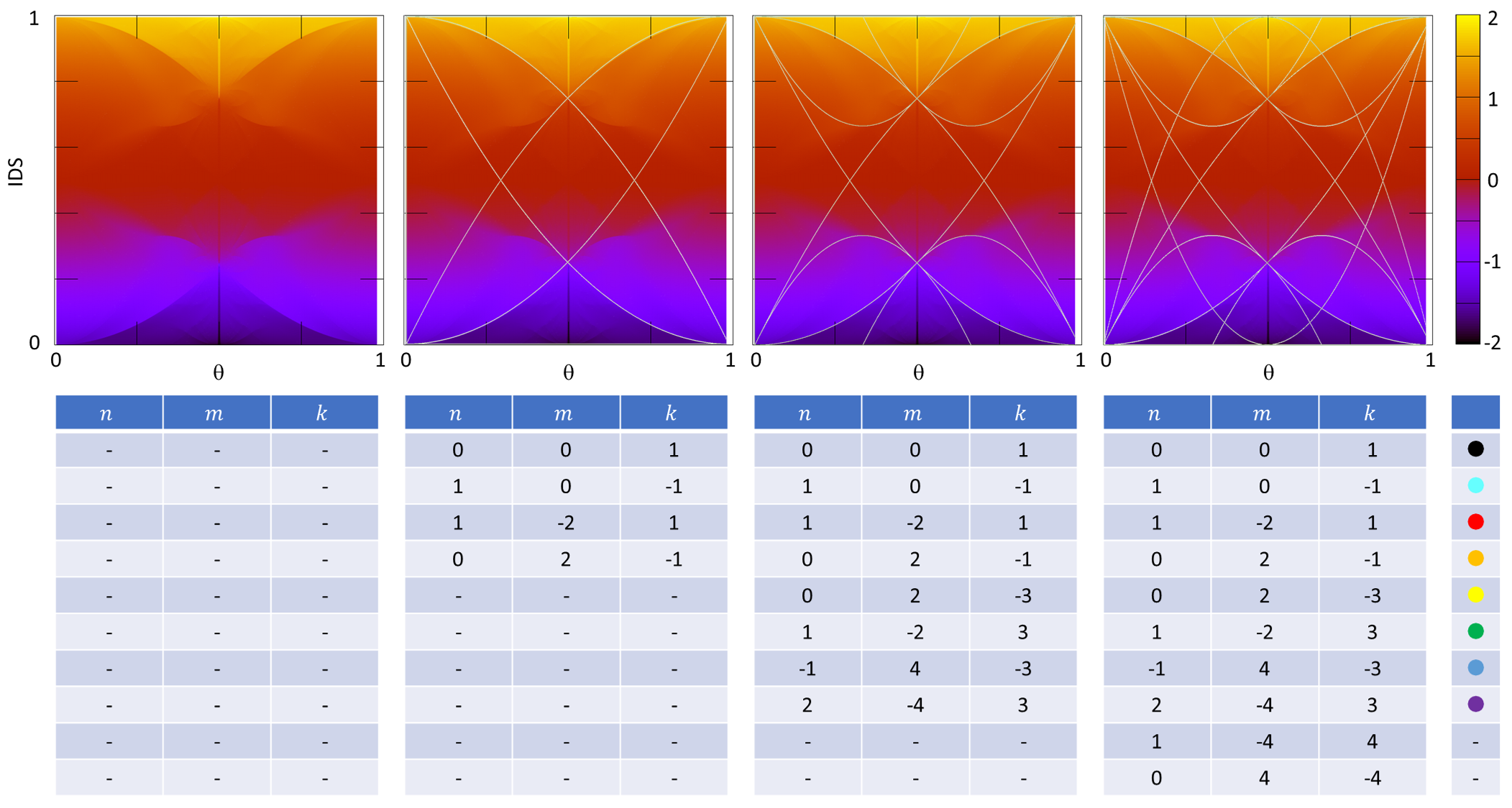}
	\caption{\small Numerical IDS as computed from the spectral butterfly reported in Fig.\ref{Fig:M2JZ0Spec} for the $M=2$ sector and $J_z=0$. The K-theoretic IDS values from Eq.~\eqref{Eq:IDS2Pred}, shown as light colored curves, are matched with the numerical IDS values inside the gaps marked in Fig.\ref{Fig:M2JZ0Spec}, identified here by the abrupt changes in the color map. The matching progresses in the order of the gap sizes. The tables list the gap labels $(n,m,k)$ from Eq.~\eqref{Eq:IDS2Pred} as well as the corresponding gaps. }
	\label{Fig:GapLabels_M2JZ0}
\end{figure*}

In Fig.~\ref{Fig:GapLabels_M1JZ0}, we report the IDS for Hamiltonian $H_1$, which has been directly computed from the spectrum ${\rm Spec}(H_1)$ reported in Fig.~\ref{Fig:M1JZ0Spec}, using the formula:
\begin{equation}\label{Eq:IDSN0}
{\rm IDS}(E) = \frac{ \big |{\rm Spec}(H_1) \cap (-\infty,E] \big | }{|\Ll|}.
\end{equation}
For lattice sizes large enough, this expression is equivalent to Eq.~\eqref{Eq:IDS1} for $d=1$.
 The IDS is represented as a function of $\theta$ and energy, with the latter on the axis coming out of page. For visualization, the energy values are encoded in the color map and the striking features seen throughout this color map are the sudden changes in color, which occur along straight lines. These sudden changes in color correspond to the spectral gaps. Indeed, since the IDS remains constant as the energy is varied inside a spectral gap, the 3-dimensional graph in Fig.~\ref{Fig:GapLabels_M1JZ0} must be aligned with the axis coming out of the page. When viewed from the top, this variation of the graph is hidden to the eye and the only thing we see is a sudden change of color. Let us point out that the larger the gap the stronger the sudden change in color. This simple phenomenon enables us to determine the numerical values of the IDS inside the prominent spectral gaps marked in Fig.~\ref{Fig:M1JZ0Spec}. Indeed, Eq.~\eqref{Eq:MainIDS} predicts the following IDS values inside the gaps:
\begin{equation}\label{Eq:IDS1Pred}
{\rm IDS}(G) \in \big \{ n + m \theta, \ n,m \in \ZM\big \} \cap [0,1],
\end{equation}
which are all linear dependencies w.r.t. $\theta$ with integer coefficients. In Fig.~\ref{Fig:GapLabels_M1JZ0}, we show how the features seen in the numerically computed IDS align with these predictions. For this, the pair ${\rm IDS}=\theta$ and ${\rm IDS} = 1-\theta$ of predicted values are laid in the second panel over the numerically computed IDS and the strongest features are identified. Additional predicted IDS values are laid in the next panel and the remaining strongest features are again identified, and similarly for the last panel. Our conclusion is that every single feature seen in the numerically computed IDS can be explained and matched by the prediction in Eq.~\eqref{Eq:IDS1Pred} derived from the K-theory of the non-commutative 2-torus.

\vspace{0.2cm}

The process explained in Fig.~\ref{Fig:GapLabels_M1JZ0} enabled us to determined the K-theoretic labels attached to the spectral gaps. They are reported in the tables in Fig.~\ref{Fig:GapLabels_M1JZ0} for the gaps marked in Fig.~\ref{Fig:M1JZ0Spec}. Since the $m$-coefficient coincides with the first Chern number, any gap carrying a non-zero $m$-label should display $m$ topological edge modes. This bulk-boundary correspondence is well understood and it is indeed confirmed by Fig.~\ref{Fig:M1JZ0Phi} and the gap labels mapped in Fig.~\ref{Fig:GapLabels_M1JZ0}. The simulations in Fig.~\ref{Fig:M1JZ0Phi} were carried with open boundary conditions, hence, the chain displays two edges and this is why the number of topological edge bands are doubled in Fig.~\ref{Fig:M1JZ0Phi}.

\vspace{0.2cm}

It will be useful for the following section to explain the $\varphi$-dependence of the spectra in the algebraic framework advocated here. For this, let $H_1(0)$ be the spin Hamiltonian corresponding to $\varphi=0$, as projected onto the $M=1$ sector. It has an expansion:
\begin{equation}
H_1(0)=\sum_{\bm q \in \ZM^2} a_{\bm q} \, U^{q_1}T^{q_2},
\end{equation}
where, for the sake of the argument, we included further neighbor couplings ({\it i.e.} $q_2$ is not restricted to just $\pm 1$). Now, to obtain the expansion of $H_1(\varphi)$ for $\varphi \neq 0$, we need to insert the factor $e^{\imath 2 \pi \varphi}$, which was taken out in \eqref{Eq:U}:
\begin{equation}
U \mapsto e^{\imath 2 \pi \varphi} U,
\end{equation}
leading to:
\begin{equation}
H_1(\varphi)=\sum_{\bm q \in \ZM^2} a_{\bm q} \, e^{\imath 2q_1 \pi \varphi}U^{q_1}T^{q_2}.
\end{equation}
When $\theta$ is irrational, all $H_1(\varphi)$, $\varphi \in \RM/\ZM$, are unitarily equivalent. For example, this is why the bulk spectrum in Fig.~\ref{Fig:M1JZ0Phi} lacks any dependence on $\varphi$. One can also convince oneself that the situation is quite different when $\theta$ is rational. On the other hand, the Hamiltonians $\widehat H_1(\varphi)$ for a semi-infinite spin-chain are no longer unitarily equivalent, regardless of the rational or irrational character of $\theta$. This is why the boundary spectrum, highlighted in red in Fig.~\ref{Fig:M1JZ0Phi}, displays a dispersion with $\varphi$.

\begin{figure}[t!]
	\includegraphics[width=\linewidth]{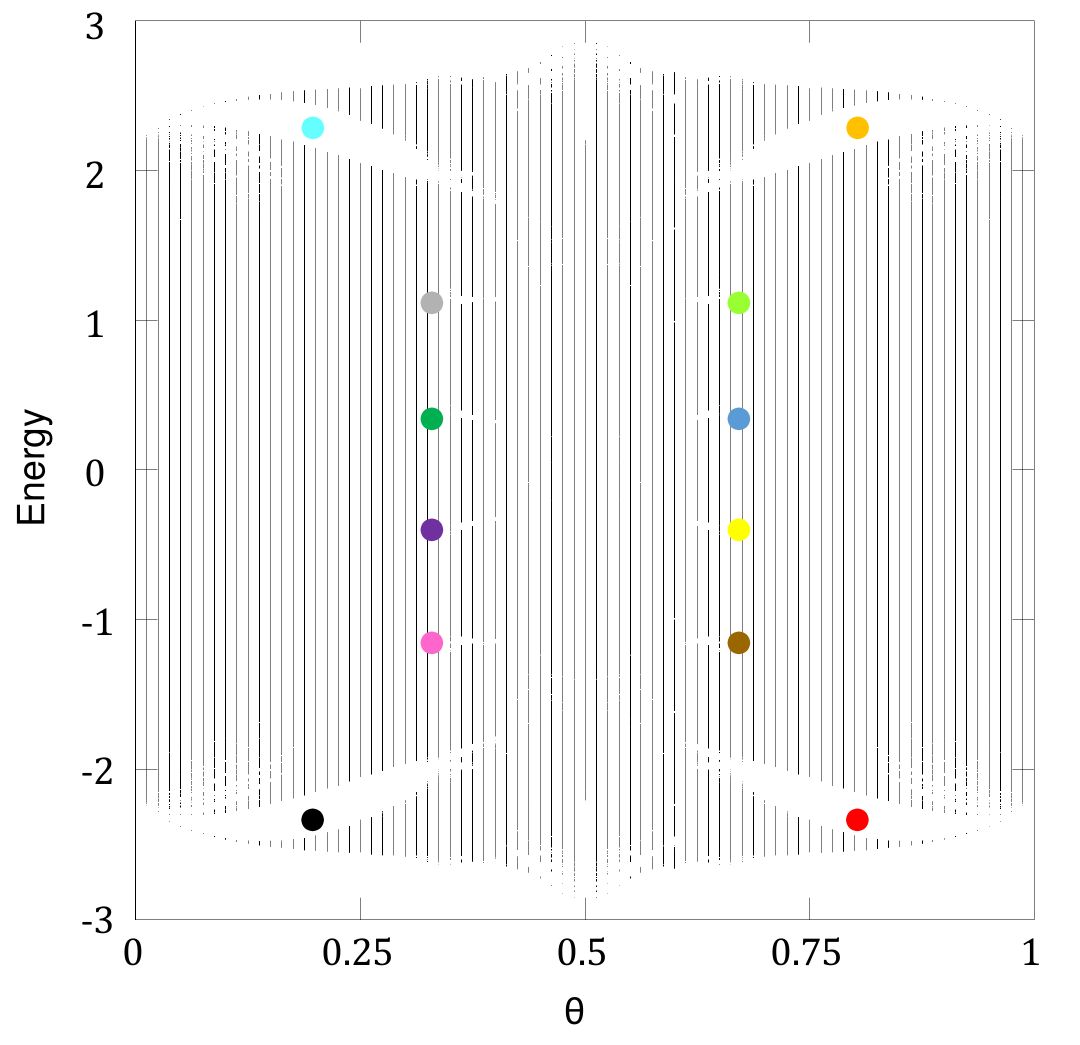}
	\caption{\small Energy spectrum of the Hamiltonian~\eqref{Eq:GenHam} as a function of $\theta$, for the $M=3$ sector and $J_z=0$. The simulation was completed for a chain with $|\Ll|= 81$ and range of the parameter $\theta$ has been sampled at rational values $\theta_n = \frac{n}{|\Ll|}$ to facilitate the closed boundary conditions. Twelve spectral gaps are identified and color-coded for future references.}
	\label{Fig:M3JZ0Spec}
\end{figure} 

\subsection{The $M=2$ Sector}
\label{Sec:M2NC}

In Fig.~\ref{Fig:M2JZ0Spec}, we report the spectrum of the spin Hamiltonian \eqref{Eq:GenHam} as a function of parameter $\theta$ from Eq.~\eqref{Eq:PattAlg}, computed inside the sector $M=2$ and with closed boundary conditions. Figure~\ref{Fig:M2JZ0Phi} reports the spectrum of the same Hamiltonian as function of parameter $\varphi$ from Eq.~\eqref{Eq:PattAlg}, computed at fixed $\theta$ and with open boundary conditions. As one can see in Fig.~\ref{Fig:M2JZ0Spec}, the fractal nature of the bulk spectrum is still apparent and the chiral edge bands are still present in Fig.~\ref{Fig:M2JZ0Phi}. This subsection is devoted to resolving the structure of the bulk spectrum, determining the K-theoretic labels associated with the spectral gaps and formulating the bulk-boundary correspondence principle, which quantitatively explains the observations in Fig.~\ref{Fig:M2JZ0Phi}.

\begin{figure}[t!]
	\includegraphics[width=\linewidth]{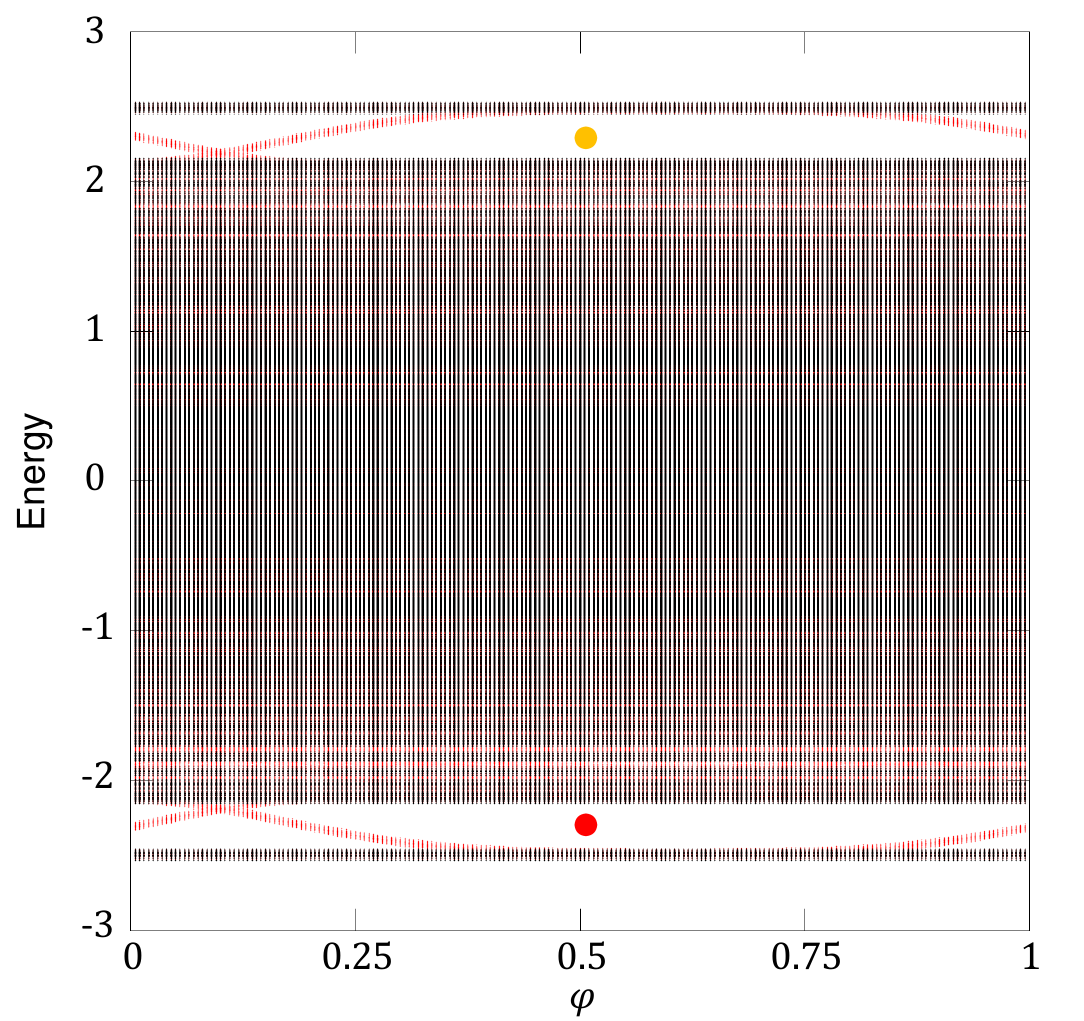}
	\caption{\small Energy spectrum of Hamiltonian \eqref{Eq:GenHam} inside the $M=3$ sector and with open boundary condition, plotted as a function of parameter $\varphi$ from Eq.~\eqref{Eq:PattAlg}. The topological boundary spectrum is highlighted in red and the spectrum with closed boundary conditions is overlaid in black color. The computation was performed with $J_z=0$, $\theta=\frac{1+\sqrt{2}}{3}$ and a chain size of 41. The colored dots mark the same gaps as in Fig.~\ref{Fig:M3JZ0Spec}.}
	\label{Fig:M3JZ0Phi}
\end{figure}

\vspace{0.2cm}

We will work directly with the fermionic representation \eqref{Eq:FHam} of the model, which needs to be projected on the 2-particle anti-symmetric Fock space $\Ff_2^{(-)} = \Hh_1 \wedge \Hh_1$ spanned by vectors of the form
\begin{equation}
\tfrac{1}{\sqrt{2}}(|n\rangle \otimes | m \rangle- |m \rangle \otimes | n \rangle) \big ), \ n,m\in \ZM.
\end{equation} 
Since the Hamiltonian is quadratic, this restriction is simply given by
\begin{equation}
H_2 = H_1 \otimes I + I \otimes H_1.
\end{equation}

\vspace{0.2cm}

\begin{figure*}[t!]
	\includegraphics[width=\linewidth]{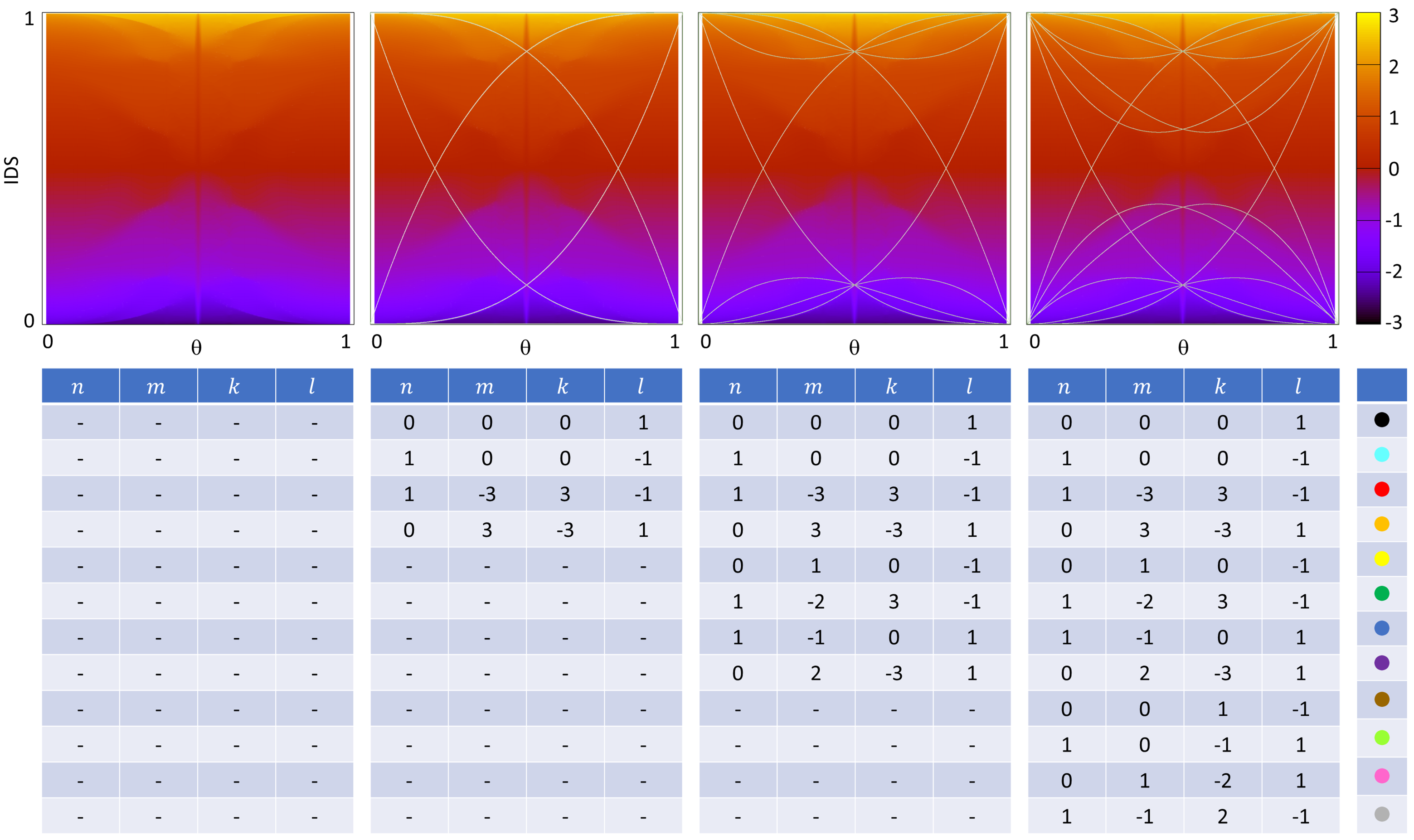}
	\caption{\small Numerical IDS as computed from the spectral butterfly reported in Fig.\ref{Fig:M3JZ0Spec} for the $M=3$ sector and $J_z=0$. The K-theoretic IDS values from Eq.~\eqref{Eq:IDS3Pred}, shown as light colored curves, are matched with the numerical IDS values inside the gaps marked in Fig.\ref{Fig:M3JZ0Spec}, identified by the abrupt changes in the color map. The matching progresses in the order of the gap sizes. The tables list the values of the integer parameters $(n,m,k,l)$ from Eq.~\eqref{Eq:IDS3Pred} as well as the corresponding gaps.}
	\label{Fig:GapLabels_M3JZ0}
\end{figure*}

A key point in our strategy is to view this Hamiltonian as acting on the full 2-particle Fock space $\Ff_2$ spanned by $|n\rangle \otimes | m \rangle$, $\ n,m\in \ZM$. If we do so, then $H_2$ belongs to the algebra generated by just four elements
\begin{equation}\label{Eq:M2Alg}
C^\ast(U \otimes I, T \otimes I, I \otimes U, I \otimes T),
\end{equation}
which can be straightforwardly shown to be the non-commutative 4-torus. Indeed, let $V_i$, $i=\overline{1,4}$, be the operators appearing in Eq.~\eqref{Eq:M2Alg}, respecting that order. Then the following commutation relations descend directly from Eq.~\eqref{Eq:Co1}:
\begin{equation}
V_1 V_2 = e^{\imath 2 \pi \theta} V_2 V_1, \quad V_3 V_4 = e^{\imath 2 \pi \theta} V_4 V_3, 
\end{equation}
and for all the remaining cases, $V_i V_j = V_j V_i$. As such, we are dealing with the non-commutative 4-torus $\Aa_{\Theta_2}$, with the $\theta$-matrix:
\begin{equation}\label{Eq:Theta2}
\Theta_2 = \begin{pmatrix} 0 & \theta & 0 & 0 \\
-\theta & 0 & 0 & 0 \\
0 & 0& 0 & \theta \\
0 & 0 & -\theta & 0 \end{pmatrix}.
\end{equation}
Let us specify that the $V_i$ generators do not preserve the anti-symmetric Fock space, while $H_2$ obviously does. So $H_2$ is generated from the symmetrized version of $\Aa_2$ but, unfortunately, that algebra does not accept a finite number of generators and relations. This is the main reason we worked with a larger algebra that, nevertheless, generates all possible $H_2$ Hamiltonians over the pattern $\Ll$. 

\vspace{0.2cm}

We computed the IDS corresponding to the spectrum ${\rm Spec}(H_2)$ reported in Fig.~\ref{Fig:M2JZ0Spec} using the following formula:
\begin{equation}\label{Eq:IDSN1}
{\rm IDS}(E) = \frac{ \big |{\rm Spec}(H_2) \cap (-\infty,E] \big | }{|\Ll|^2},
\end{equation}
and the results are reported in Fig.~\ref{Fig:GapLabels_M2JZ0}. As one can see, the lines where the color changes abruptly are now curved instead of being linear. Using the same argument as before, these lines are identified with the values of the IDS inside the spectral gaps. For those cases, Eq.~\ref{Eq:IDSN1} can be shown to be equivalent to $\Tt(p_G)$ and the predictions spelled in \eqref{Eq:MainIDS} apply. With the $\Theta$ from \eqref{Eq:Theta2}, these predictions translate to
\begin{equation}\label{Eq:IDS2Pred}
{\rm IDS}(G) \in \big \{n + m \theta + k \theta^2, \ n,m,k \in \ZM \big \} \cap [0,1].
\end{equation}
Using the same strategy as for the case of $M=1$, we demonstrate in Fig.~\ref{Fig:GapLabels_M2JZ0} that the curves mentioned above match most of the features seen in the numerically computed IDS. In fact, further investigations, which are not reported here, convinced us that we can match all the features seen in the numerical IDS. The process used in Fig.~\ref{Fig:GapLabels_M2JZ0} enabled us to identify the K-theoretic labels for the spectral gaps, which are reported in the tables of Fig.~\ref{Fig:GapLabels_M2JZ0}. As one can see, the top index $k$ is non-zero for all identified spectral gaps.

\vspace{0.2cm}

\begin{figure*}[t!]
	\includegraphics[width=\linewidth]{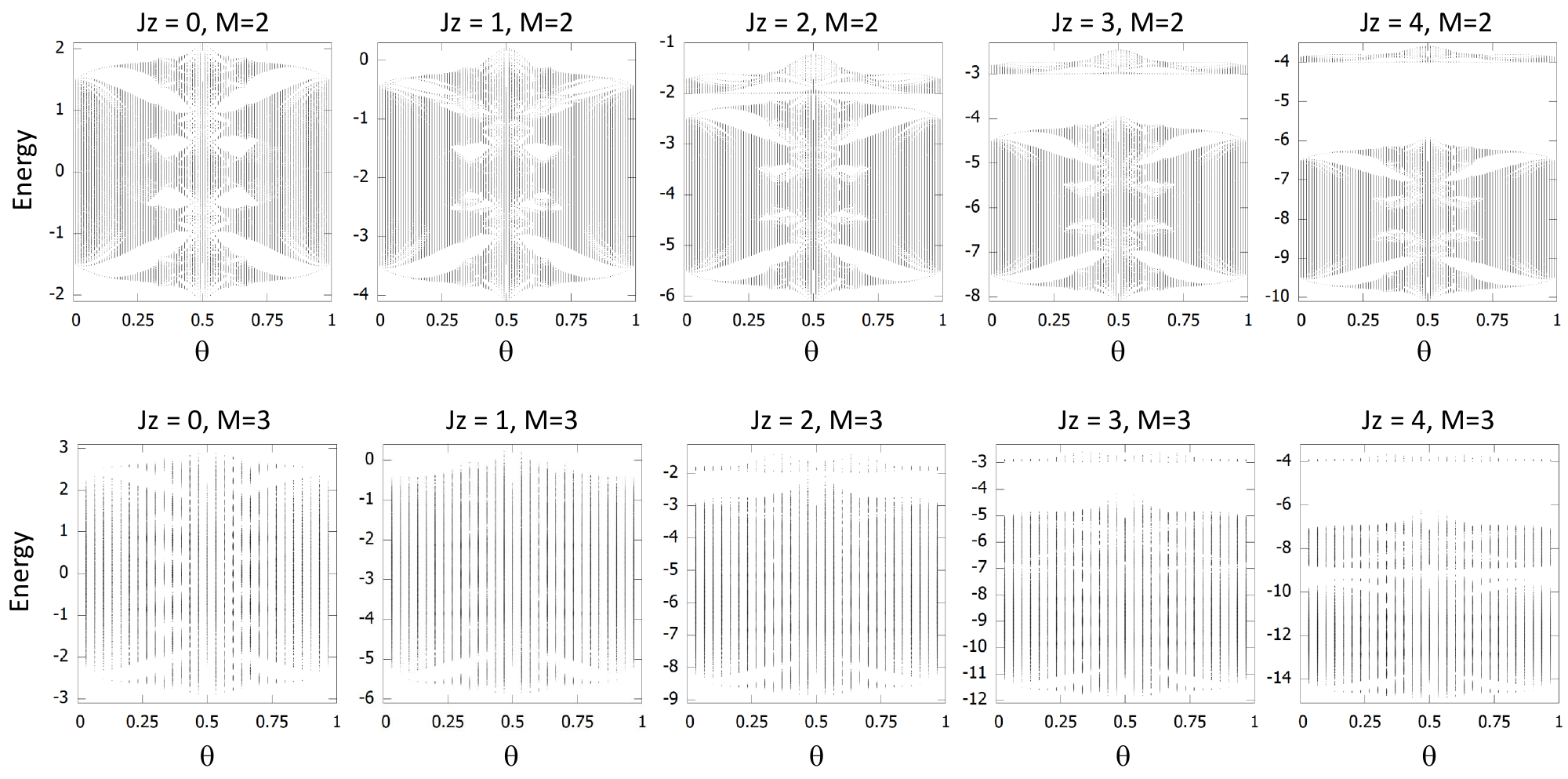}
	\caption{\small Energy spectrum of the Hamiltonian \eqref{Eq:GenHam} with closed boundary conditions, plotted as a function of parameter $\theta$ from Eq.~\eqref{Eq:PattAlg} and for the specified values of interaction strength $J_z$. The top row corresponds to the $M=2$ sector and for these simulations $|\Ll|=101$. The bottom row corresponds to the $M=3$ sector and for these simulations $|\Ll|=31$. In all panels, the energies are referenced from $\frac{|\Ll|}{4}J_z$.}
	\label{Fig:JzSpecEvolution}
\end{figure*}

We now turn our attention to the bulk-boundary principle. In Fig.~\ref{Fig:M2JZ0Phi}, one can  see bundles of chiral bands of both positive and negative slope. Our next goal is to explain, quantitatively, the structure of these chiral bands. The standard bulk-boundary correspondence for the non-commutative 4-torus was worked out in details in \cite{ProdanSpringer2016}. However, for the spin system, the bulk-boundary correspondence is more complicated but also more interesting. Indeed, note that:
\begin{equation}\label{Eq:XP1}
H_2(\varphi) = H_1(\varphi) \otimes I + I \otimes H_1(\varphi).
\end{equation}
If we could engineer the Hamiltonian
\begin{equation}\label{Eq:Phi12Ham}
H_2(\varphi_1,\varphi_2) = H_1(\varphi_1) \otimes I + I \otimes H_1(\varphi_2),
\end{equation} with independent control on $\varphi_1$ and $\varphi_2$, then Eq.~\eqref{Eq:XP1} would have been entirely equivalent to a topological insulator from class A in $d=4$. Indeed, $\varphi_{1,2}$ will represent two virtual momenta and the fermionic system will have 2 physical and 2 virtual dimensions. However, the two identical fermions experience the same underlying pattern so the two virtual momenta are bound to be the same. The consequence is that we will not be able to explore the full dispersion of the boundary states but only the diagonal sector $\varphi_1=\varphi_2$. Furthermore, when a physical edge is imposed on the spin chain, both virtual fermions experience the boundary. In other words, the Hamiltonian with a boundary becomes
\begin{equation}
\widehat H_2(\varphi) = \widehat H_1(\varphi) \otimes I + I \otimes \widehat H_1(\varphi),
\end{equation}
where $\widehat H_1(\varphi)$ is the Hamiltonian mentioned in Sec.~\ref{Sec:M1Jz0}, and the physical edge of the 1-dimensional spin-chain becomes a hinge for the virtual 4-dimensional system. The boundary states seen in Fig.~\ref{Fig:M2JZ0Phi} are not related to the hinge states studied in \cite{BenalcazarPRB2017}, which are stabilized by a point symmetry, neither to the ones studied in \cite{HayashiCMP2018}, which require gapped boundaries. Corner-following states in a hinged geometry were studied in \cite{ThiangArxiv2019}, but the boundary states observed in Fig.~\ref{Fig:M2JZ0Phi} are not related to such states either, because  Fig.~\ref{Fig:M2JZ0Phi} is about the dispersion of the boundary modes with respect to momenta in planes perpendicular to the boundaries.

\vspace{0.2cm}

In the following, we adapt the bulk-boundary formalism from \cite{ProdanSpringer2016} to the new setting, with the goal of deriving a quantitatively precise bulk-boundary principle for the spin chain. First, we will establish the bulk-boundary correspondence for the pair $(H_2,\widehat H_2)$ on the full Fock space $\Ff_2$ and we will project onto the anti-symmetrized sector $\Ff_2^{(-1)}$ at the end. We start by noticing that, topologically, the diagonal path inside the $(\varphi_1,\varphi_2)$-torus of Eq.~\eqref{Eq:Phi12Ham} is equivalent to the concatenation of two paths
\begin{align}
\{(\varphi,\varphi),\ \varphi\in [0,1] \} \simeq & \{(\varphi_1,0), \ \varphi_1 \in [0,1]\} \\ \nonumber 
& \quad \cup  \{(1,\varphi_2), \ \varphi_2 \in [0,1]\}.
\end{align}
Since, the net number $N$ \cite{Footnote1} of chiral bands traversing the bulk gap $G$ does not change under such deformations, we reduced the problem to counting the chiral modes of $\widehat H(\varphi_1,\varphi_2)$ emerged when varying $\varphi_1$ with $\varphi_2$ kept constant plus the ones emerged when varying $\varphi_2$ with $\varphi_1$ kept constant. Due to the particular form of the Hamiltonian \eqref{Eq:Phi12Ham}, this count is given by
\begin{equation}
N/|\Ll| = {\rm Ch}_{\{1,2\}}(P_G) + {\rm Ch}_{\{3,4\}}(P_G),
\end{equation}
which is expected to hold for large $|\Ll|$. Note that we excluded ${\rm Ch}_{\{2,3\}}(P_G)$ from the count, on the basis that an edge in the third direction ({\it i.e.} on the second fermion) will produce disperseless boundary modes with respect to $\varphi_1$. For similar reasons, we have also excluded ${\rm Ch}_{\{1,4\}}(P_G)$ from the count. Finally, the projection onto $\Ff_2^{(-1)}$ should reduce this count by a factor of 2.

\begin{figure*}[t!]
	\includegraphics[width=\linewidth]{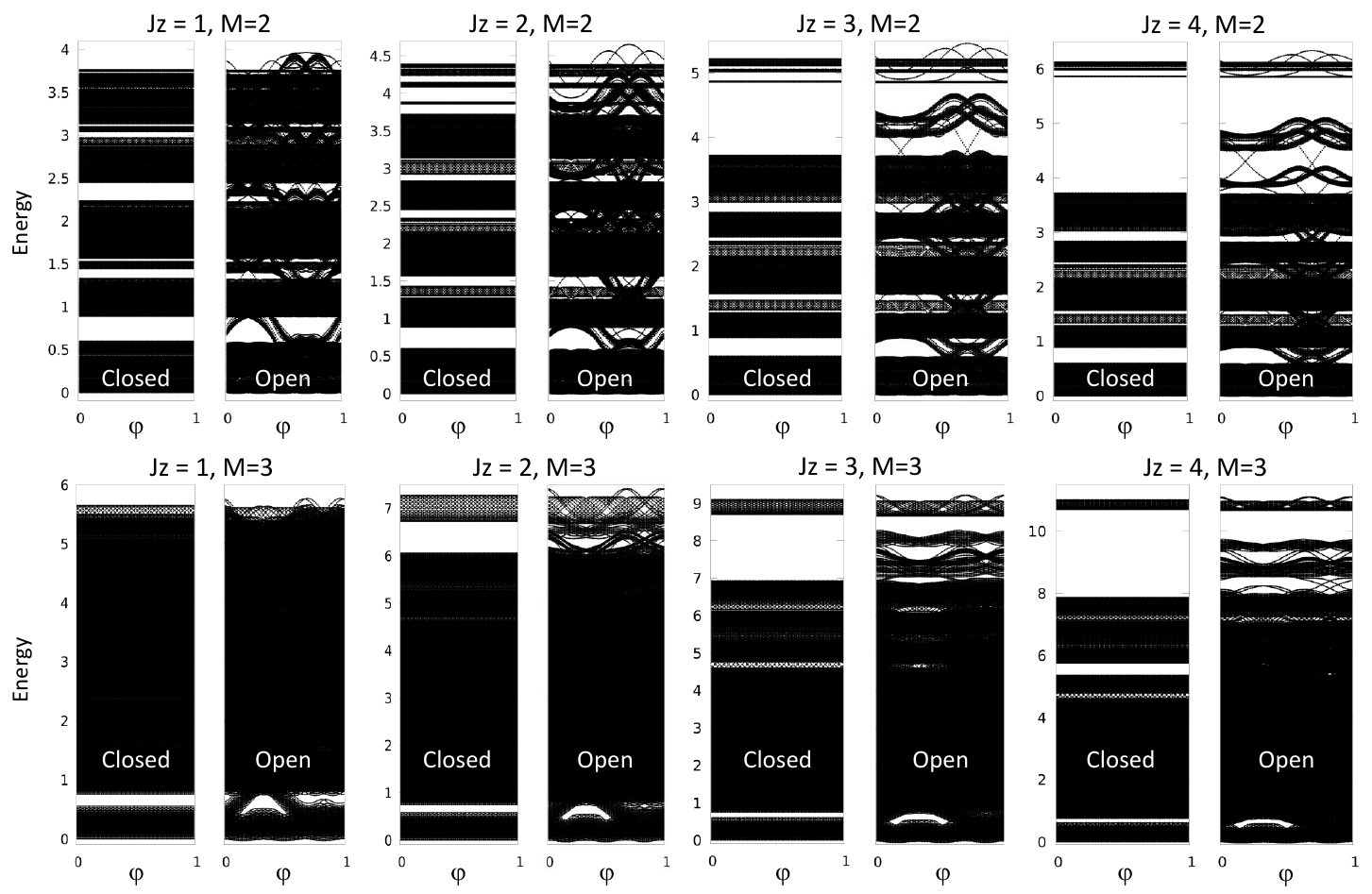}
	\caption{\small Energy spectrum of Hamiltonian \eqref{Eq:GenHam} with closed/open boundary conditions, plotted as a function of parameter $\varphi$ from Eq.~\eqref{Eq:PattAlg} and for specified values of interaction strength $J_z$. The top row corresponds to the $M=2$ sector and for these simulations $\theta=\frac{1+\sqrt{2}}{4}$ and $|\Ll_L| = 59$, while the bottom row corresponds to the $M=3$ sector and for these simulations $\theta=\frac{1+\sqrt{2}}{7}$ and $|\Ll_L|=29$. To facilitate comparisons, the energies are referenced from the bottom of the spectra in all panels.}
	\label{PhiDependenceInter}
\end{figure*}

\vspace{0.2cm}

Using \eqref{Eq:ChernVal}, we now can state a quantitative bulk-boundary principle:
\begin{align}
\lim_{|\Ll|\rightarrow \infty}\frac{N}{|\Ll|} = & \tfrac{1}{2} \Big [n_{\{1,2\}} + n_{\{3,4\}}+ \\ \nonumber 
& \quad n_{\{1,2,3,4\}} \big ({\rm Pf}(\Theta_{\{1,2\}})+ {\rm Pf}(\Theta_{\{3,4\}}) \big ) \big ],
\end{align}
which further simplifies if we use the relation between the $(n,m)$ gap labels and the coefficients $n_J$ in Sec.~\ref{Ch:KTh}:
\begin{equation}\label{Eq:BBM2}
\lim_{|\Ll|\rightarrow \infty}\frac{N}{|\Ll|} = \tfrac{1}{2} (m+2k \theta ).
\end{equation}

\begin{table}[b!]
  \begin{center}
    \label{Tab:Table1}
    \begin{tabular}{c|c|c} 
    \hline 
    \ Gap \ & $N/{|\Ll|}$ by direct count  & Prediction from  Eq.~\eqref{Eq:BBM2} \\
      \hline \hline
      \textcolor{black}{\large $\bullet$} & 0.3548 & 0.3964\\
      \textcolor{yellow}{\large $\bullet$} & -0.193548 & -0.18934\\
      \textcolor{ForestGreen}{\large $\bullet$} & 0.193548 & 0.18934\\
      \textcolor{Cyan}{\large $\bullet$} & -0.3548 & -0.3964\\
      \hline
    \end{tabular}
    \caption{Bulk-boundary principle for $M=2$ sector and $J_z=0$, tested for a chain with open boundary conditions, $|\Ll|=31$ and $\theta =1-\frac{1+\sqrt{2}}{4}$.}
  \end{center}
\end{table}

\vspace{0.2cm}

In Table~I, we supply a comparison between the left side of Eq.~\eqref{Eq:BBM2}, as computed by a direct count of the edge modes, and the righ side of Eq.~\eqref{Eq:BBM2}, as computed from the gap labels listed in Fig.~\ref{Fig:GapLabels_M2JZ0}. The matching between the two is remarkable, given the relatively small size of the system \cite{Footnote2}.

\subsection{The $M=3$ Sector}
\label{Sec:M3NC}

The spectrum of the spin Hamiltonian \eqref{Eq:GenHam} inside the $M=3$ sector and with closed boundary conditions is reported in Fig.~\ref{Fig:M3JZ0Spec} as a function of $\theta$. As one can see, there is still strong evidence of fractality, but the number of open gaps is much reduced when compared with the previous cases. Furthermore, when we open the boundary conditions, chiral edge modes are again observed in Fig.~\ref{Fig:M3JZ0Phi} and clear features in the numerical IDS reported in Fig.~\ref{Fig:GapLabels_M3JZ0} can be again identified. By following closely the analysis for the $M=2$, we show again that the spectral features can be completely explained by the K-theory. Furthermore, it will be shown that the gaps seen in the spectrum carry strong topological numbers and that they display non-trivial boundary spectrum.

\vspace{0.2cm} 

\begin{figure}[t!]
	\includegraphics[width=0.85\linewidth]{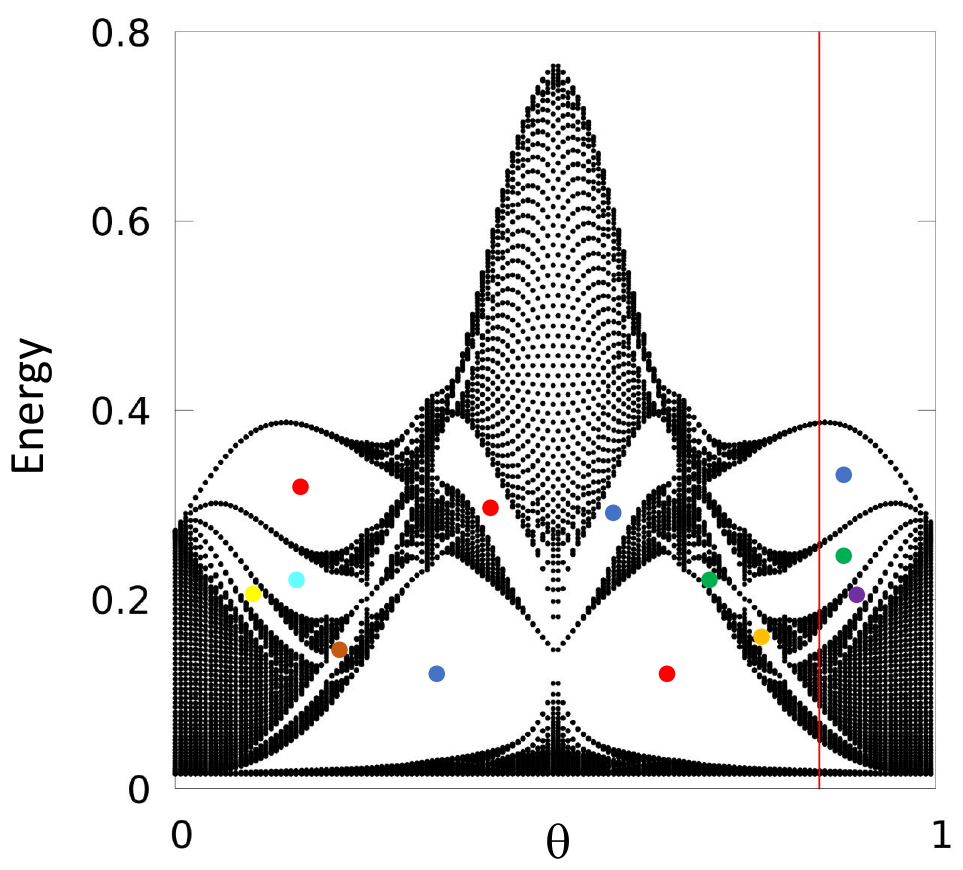}
	\caption{\small A refined representation of the top spectral butterfly identified in Fig.~\ref{Fig:JzSpecEvolution} for the $M=2$ sector. The simulation parameters are $r=0.45$, $J_z=2.1$ and $|\Ll|=151$. The red vertical line indicates the value of $\theta=\frac{2+\sqrt{5}}{5}$ where the bulk-boundary correspondence will be probed. The spectral gaps are labeled by colored dots for future reference. }
	\label{Fig:BulkSpecL151M2Jz2p1Eps0p0Top}
\end{figure}

Working directly with the fermionic representation and restricting $H_F$ from Eq.~\eqref{Eq:FHam} to the 3-particle anti-symmetric Fock space $\Ff_3^{(-)} = \Hh_1 \wedge \Hh_1\wedge \Hh_1$, the Hamiltonian becomes
\begin{equation}
H_3 = H_1 \otimes I\otimes I + I \otimes H_1\otimes I+I\otimes I\otimes H_1.
\end{equation}
Each of the terms in $H_3$ can be generated from the following set of operators:
\begin{align}
V_1 = U \otimes I \otimes I, \quad V_2 = T \otimes I \otimes I, \nonumber \\
V_3 = I \otimes U \otimes I, \quad V_4 = I \otimes T \otimes I, \\
V_5 = I \otimes I \otimes U, \quad V_6 = I \otimes I \otimes T, \nonumber
\end{align}
acting on the full 3-particle Fock space $\Ff_3$. Hence, the Hamiltonian belongs to the algebra $C^\ast\big (V_i, \ i = \overline{1,6} \big )$ generated by the $V_i$'s, which can be straightforwardly shown to be the non-commutative 6-torus $\Aa_{\Theta_3}$, with the $\theta$-matrix:
\begin{equation}\label{Eq:Theta3}
\Theta_3 = \begin{pmatrix} 0 & \theta & 0 & 0 & 0 & 0 \\
-\theta & 0 & 0 & 0 & 0 & 0 \\
0 & 0& 0 & \theta & 0 & 0 \\
0 & 0 & -\theta & 0 & 0 & 0 \\
0 & 0& 0 & 0 & 0 & \theta \\
0 & 0 & 0 & 0 & -\theta & 0
 \end{pmatrix}.
\end{equation}

\begin{figure}[t!]
	\includegraphics[width=\linewidth]{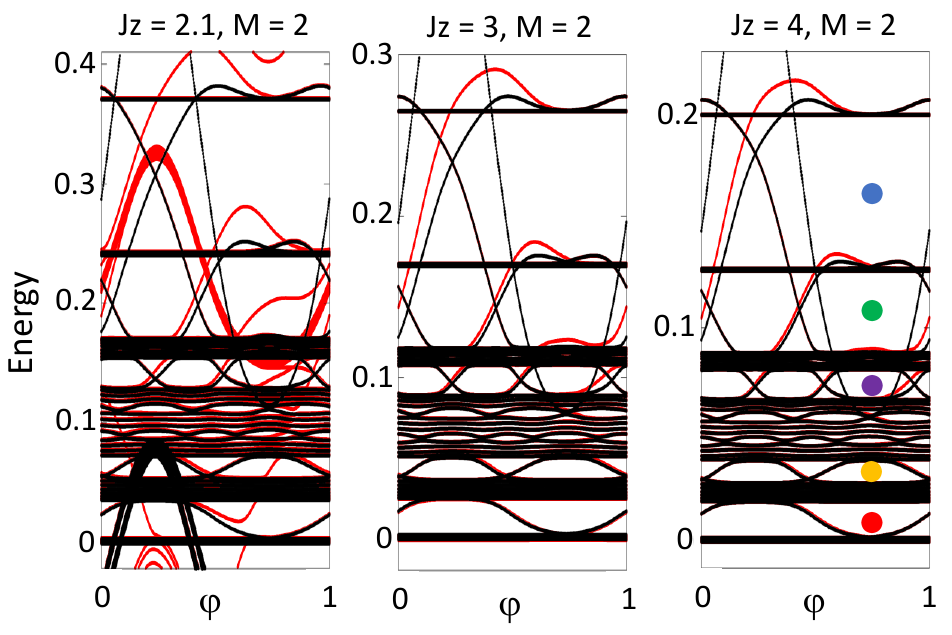}
	\caption{\small Top spectral island of Hamiltonian \eqref{Eq:GenHam} computed with open boundary condition as function of $\varphi$ at fixed $\theta=\frac{2+\sqrt{5}}{5}$ (see vertical line in Fig.~\ref{Fig:BulkSpecL151M2Jz2p1Eps0p0Top}), and chain length $|\Ll|=61$. The colored dots labeling the gaps are correlated with the ones in  Fig.~\ref{Fig:BulkSpecL151M2Jz2p1Eps0p0Top}. The spectra were computed in two ways, with (red lines) and without (black lines) a defect potential on the left edge of the chain. This enabled us to identify the chiral bands located at the left edge of the chain, which are the ones for which the black and the red simulations do not overlap. Specifically, the gaps \textcolor{blue}{\large $\bullet$}/\textcolor{green}{\large $\bullet$}/\textcolor{violet}{\large $\bullet$} display 1/2/3 positively sloped chiral bands localized at the left edge, respectively, while the gaps \textcolor{red}{\large $\bullet$}/\textcolor{orange}{\large $\bullet$} display 1/2 negatively sloped chiral bands localized at the left edge. This is in perfect agreement with the gap labels derived in Fig.~\ref{Fig:GapLabels_M2JZ2p1}.}
	\label{Fig:PhiDependenceM2Top}
\end{figure}

\vspace{0.2cm}

The IDS reported in Fig.~\ref{Fig:GapLabels_M3JZ0} was computed from the spectrum ${\rm Spec}(H_3)$ reported in Fig.~\ref{Fig:M3JZ0Spec} using the formula
\begin{equation}\label{Eq:IDSN2}
{\rm IDS}(E) = \frac{ \big |{\rm Spec}(H_3) \cap (-\infty,E] \big | }{|\Ll|^3},
\end{equation}
which can be shown again to coincide with $\Tt(p_G)$ when the energy takes values inside the spectral gap $G$. As such, the prediction from Eq.~\eqref{Eq:MainIDS} applies, which together with the $\Theta$ reported above, lead to the prediction
\begin{equation}\label{Eq:IDS3Pred}
{\rm IDS}(G) \in \{n + m \theta + k \theta^2 + l \theta^3, \ n,m,k,l \in \ZM\} \cap [0,1].
\end{equation}
In Fig.~\ref{Fig:GapLabels_M3JZ0} we demonstrate that these predictions match most of the features seen in the numerical IDS. In the process, we were able again to identify the K-theoretic labels of the spectral gaps. Interestingly, we find again that all gaps carry a non-zero top index $l$, which is equal to the top Chern number in dimension 6. As such, topological boundary spectrum is expected when the boundary conditions are opened. 

\vspace{0.2cm}

\begin{figure*}[t!]
	\includegraphics[width=\linewidth]{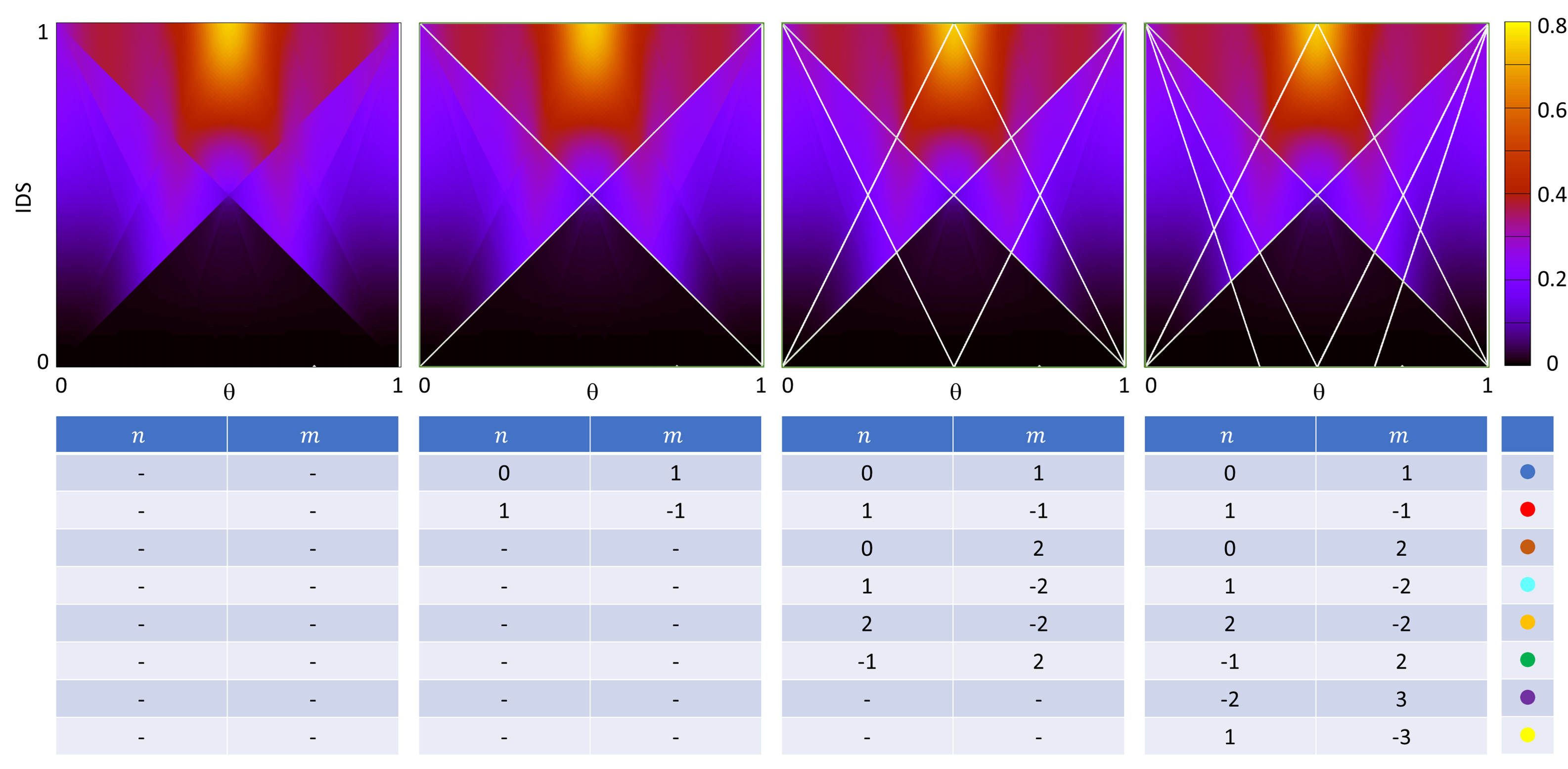}
	\caption{\small Numerical IDS as computed from the spectral butterfly reported in Fig.~\ref{Fig:BulkSpecL151M2Jz2p1Eps0p0Top} for the $M=2$ sector and $J_z=2.1$. The K-theoretic IDS values from Eq.~\eqref{Eq:IDS1Pred}, shown as light colored lines, are matched with the numerical IDS values inside the gaps marked in Fig.~\ref{Fig:BulkSpecL151M2Jz2p1Eps0p0Top}, identified by the abrupt changes in the color plot. The matching progresses in the order of the gap sizes. The tables list the values of the two integer parameters $(n,m)$ from Eq.~\eqref{Eq:IDS1Pred} as well as the corresponding gaps.}
	\label{Fig:GapLabels_M2JZ2p1}
\end{figure*}

Our next goal is to explain, quantitatively, the structure of the chiral bands in Fig.~\ref{Fig:M3JZ0Phi}. Writing out the $\varphi$-dependency, we have
\begin{equation}\label{Eq:XP2}
H_3(\varphi) = H_1(\varphi) \otimes I \otimes I + I \otimes H_1(\varphi) \otimes I + I \otimes I \otimes H_1(\varphi).
\end{equation}
As before, because the virtual fermions experience the same potential, we can only explore the diagonal sector of the fully general Hamiltonian
\begin{align}\label{Eq:H3Full}
H_3(\varphi_1,\varphi_2,\varphi_3) & = H_1(\varphi_1) \otimes I \otimes I  \\ \nonumber
& \qquad + I \otimes H_1(\varphi_2) \otimes I + I \otimes I \otimes H_1(\varphi_3).
\end{align}
Nevertheless, we observe again that, topologically, the diagonal path inside the $(\varphi_1,\varphi_2,\varphi_3)$ torus is equivalent to the concatenation of three paths
\begin{align}
\{(\varphi,\varphi,\varphi),&\ \varphi\in [0,1] \} \simeq  \{(\varphi_1,0,0), \ \varphi_1 \in [0,1]\} \\ \nonumber 
& \cup  \{(1,\varphi_2,0), \ \varphi_2 \in [0,1]\}  \cup  \{(1,1,\varphi_3), \ \varphi_3 \in [0,1]\}.
\end{align}
As for the case $M=2$, since the net number of chiral edge bands do not change under such deformations, we can use these three simpler paths to conclude that
\begin{equation}
N/|\Ll|^2 = {\rm Ch}_{1,2}(P_G)+{\rm Ch}_{3,4}(P_G)+{\rm Ch}_{5,6}(P_G).
\end{equation}
Projection onto the anti-symmetric Fock space $\Ff_3^{(-)}$ should reduce the count by a factor of $3!$. Then, using \eqref{Eq:ChernVal}, we can state a quantitative bulk-boundary principle
\begin{align}
\lim_{|\Ll|\rightarrow \infty}\frac{N}{|\Ll|^2} = & \tfrac{1}{6} \Big [n_{\{1,2\}} + n_{\{3,4\}} +  n_{\{5,6\}} + (n_{\{1,2,3,4\}}  \\ \nonumber 
&   +n_{\{1,2,5,6\}} + n_{\{3,4,5,6\}} ) 2\theta+ n_{\{1,2,3,4,5,6\}} 3\theta^2\big ].
\end{align}
which further simplifies if we use the relation between the $(m,k,l)$ gap labels and the coefficients $n_J$ in Sec.~\ref{Ch:KTh}:
\begin{equation}\label{Eq:BBM3}
\lim_{|\Ll|\rightarrow \infty}\frac{N}{|\Ll|^2} = \tfrac{1}{6} (m+2k \theta +3l\theta^2 ).
\end{equation}

\begin{table}[b!]
	\begin{center}
		\label{Tab:Table2}
		\begin{tabular}{c|c|c} 
			\hline 
			\ Gap \ & $N/{|\Ll|}$ by direct count  & Prediction from  Eq.~\eqref{Eq:BBM3} \\
			\hline
			\textcolor{YellowOrange}{\large $\bullet$} & 0.1249 & 0.1906\\
			\textcolor{Red}{\large $\bullet$} & -0.1249 & -0.1906\\
			\hline
		\end{tabular}
		\caption{Bulk-boundary principle for $M=3$ sector and $J_z=0$, tested for a chain with open boundary conditions, $|\Ll|=41$ and $\theta =\frac{1+\sqrt{2}}{3}$.}
	\end{center}
\end{table}

In Table~II, we supply a comparison between the left side of Eq.~\eqref{Eq:BBM3}, as computed by a direct count of the edge modes, and the righ side of Eq.~\eqref{Eq:BBM3}, as computed from the gap labels listed in Fig.~\ref{Fig:GapLabels_M3JZ0}. We atribute the slight difference on the slow convergence to the thermodynamic limit, which we plan to further investigate in the future.

\section{Topological Gaps: The Correlated Case}
\label{Sec:TopoGaps2}

The evolution of the spectral butterflies with the strength $J_z$ of the interaction is reported in Fig.~\ref{Fig:JzSpecEvolution} for both sectors $M=2$ and $3$. As one can see, the fractal nature of the spectrum persists and interesting islands of spectrum separate at strong $J_z$. Furthermore, as shown in Fig.~\ref{PhiDependenceInter}, when computed with open boundary conditions at a fixed $\theta$, the spectra continue to display a rich structure of chiral edge bands as the parameter $\varphi$ is varied. This section is devoted to understanding these spectra through the prism of generating algebras and their K-theories.

\begin{figure*}[t!]
	\includegraphics[width=0.9\linewidth]{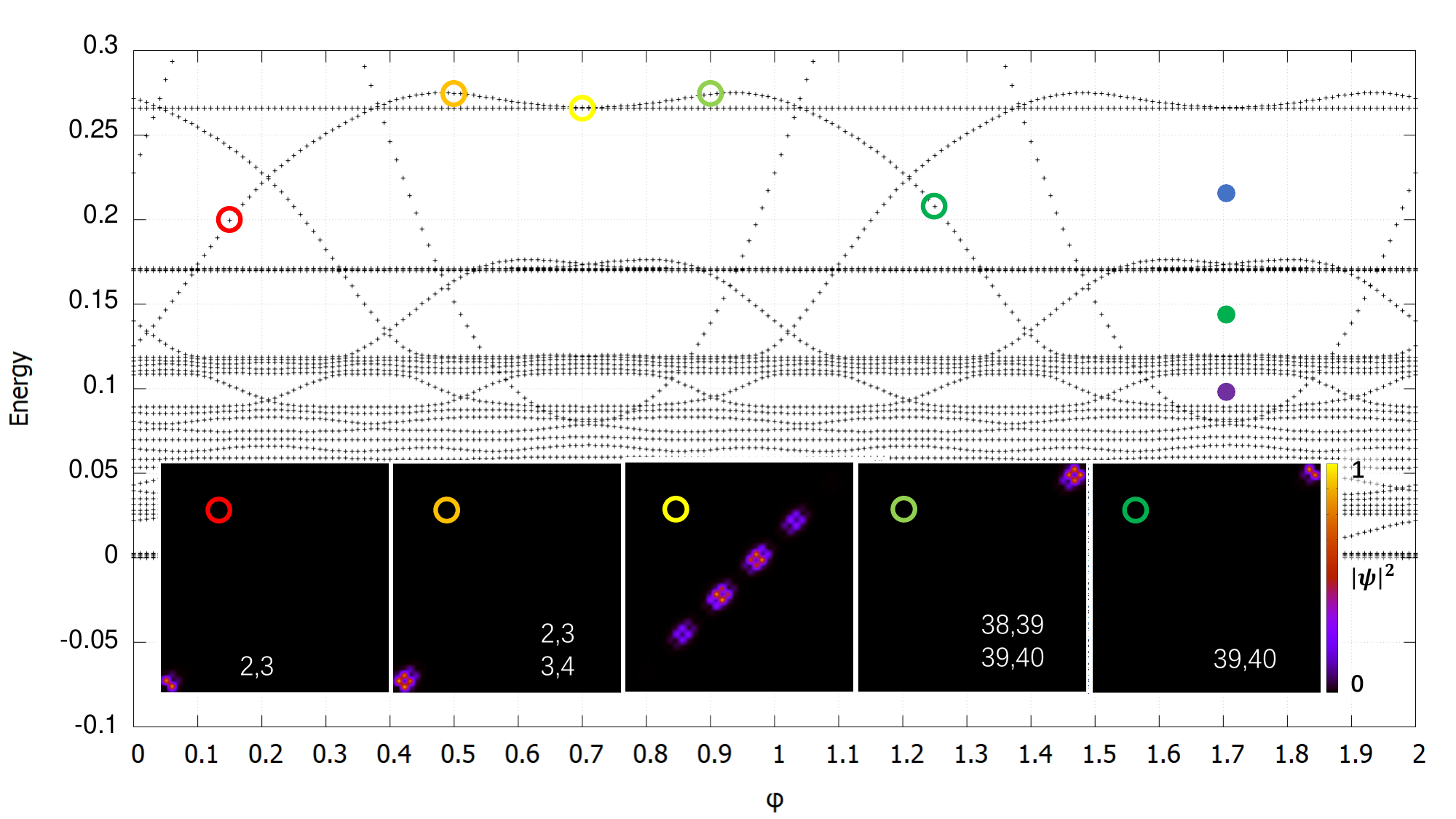}
	\caption{\small Visualization of the chiral edge modes associated to the top spectral island of the Hamiltonian~\eqref{Eq:GenHam} in the $M=2$ sector. The eigenvalues and the corresponding wave-functions are color-coded. The simulation parameters are $r=0.45$, $\theta=\frac{2+\sqrt{5}}{5}$, $J_z=3$ and $|\Ll|=41$. The intensity maps represent the probabilities for two spins at locations $n$ and $m$ along the chain to be flipped. The numbers seen in some of the panels represent the coordinates $(n,m)$ where the probabilities take significant values. The colored dots labeling the gaps are correlated with the ones in  Fig.~\ref{Fig:BulkSpecL151M2Jz2p1Eps0p0Top}.}
	\label{EdgeModeM2Jz3EM1}
\end{figure*}

\subsection{The $M=2$ sector}
\label{Sec:M2Corr}

We will take first a closer look at the case $J_z = 2.1$, which is the interaction strength where the spectral islands are already separated in Fig.~\ref{Fig:JzSpecEvolution}. The top spectral butterfly, computed with closed boundary conditions, is shown in more details in Fig.~\ref{Fig:BulkSpecL151M2Jz2p1Eps0p0Top}, with the energy referenced from the bottom of this top spectral island. When the boundary condition is opened and the spectrum is computed as function of $\varphi$ at fixed $\theta$, clear chiral bands develop as shown in Fig.~\ref{Fig:PhiDependenceM2Top}. Furthermore, the IDS corresponding to spectral butterfly in Fig.~\ref{Fig:BulkSpecL151M2Jz2p1Eps0p0Top}, shown in Fig.~\ref{Fig:GapLabels_M2JZ2p1}, displays the same straight lines seen in the non-correlated ($J_z=0$) $M=1$ case. It becomes evident that in Fig.~\ref{Fig:BulkSpecL151M2Jz2p1Eps0p0Top} we are seeing a highly distorted but nevertheless a Hofstadter butterfly, hence, we are dealing again with the non-commutative 2-torus. The first part of the section is devoted to understanding this empirical observation.

\vspace{0.2cm}

When restricted to the $M=2$ sector, the interaction potential in \eqref{Eq:GenHam}, reduces to:
\begin{equation}\label{Eq:IntM2}
J_z \, \sum_{n} S^{z}_n S^z_{n+1} \mapsto \Lambda_1 P_1 + \Lambda_2 (I-P_1),
\end{equation}
where $P_1$ is the projection onto the sub-space spanned by the states
\begin{equation}
|\psi_{n}\rangle = \tfrac{1}{\sqrt{2}}\big (|n\rangle \otimes |n+1\rangle - |n+1\rangle \otimes |n\rangle \big ) \in \Ff_2^{(-)},
\end{equation}
with $n$ running over all integer values, and
\begin{equation}
\Lambda_1 = -J_z, \quad \Lambda_2 = -2J_z,
\end{equation}
when the energy is referenced from $\frac{|\Ll|}{4}J_z$, as it was already done in  Fig.~\ref{Fig:JzSpecEvolution}. In the limit of strong interaction, the potential \eqref{Eq:IntM2} dominates and, as such, it dictates the global structure of the spectrum. In particular, it separates the energy spectrum in two spectral islands, as already seen in Fig.~\ref{Fig:JzSpecEvolution}. 
For example, we have verified that the spectral gap $G_0$ that separates these islands becomes assymptotically equal to $J_z$ when $J_z \rightarrow \infty$.

\vspace{0.2cm}

\begin{figure}[t!]
	\includegraphics[width=0.9\linewidth]{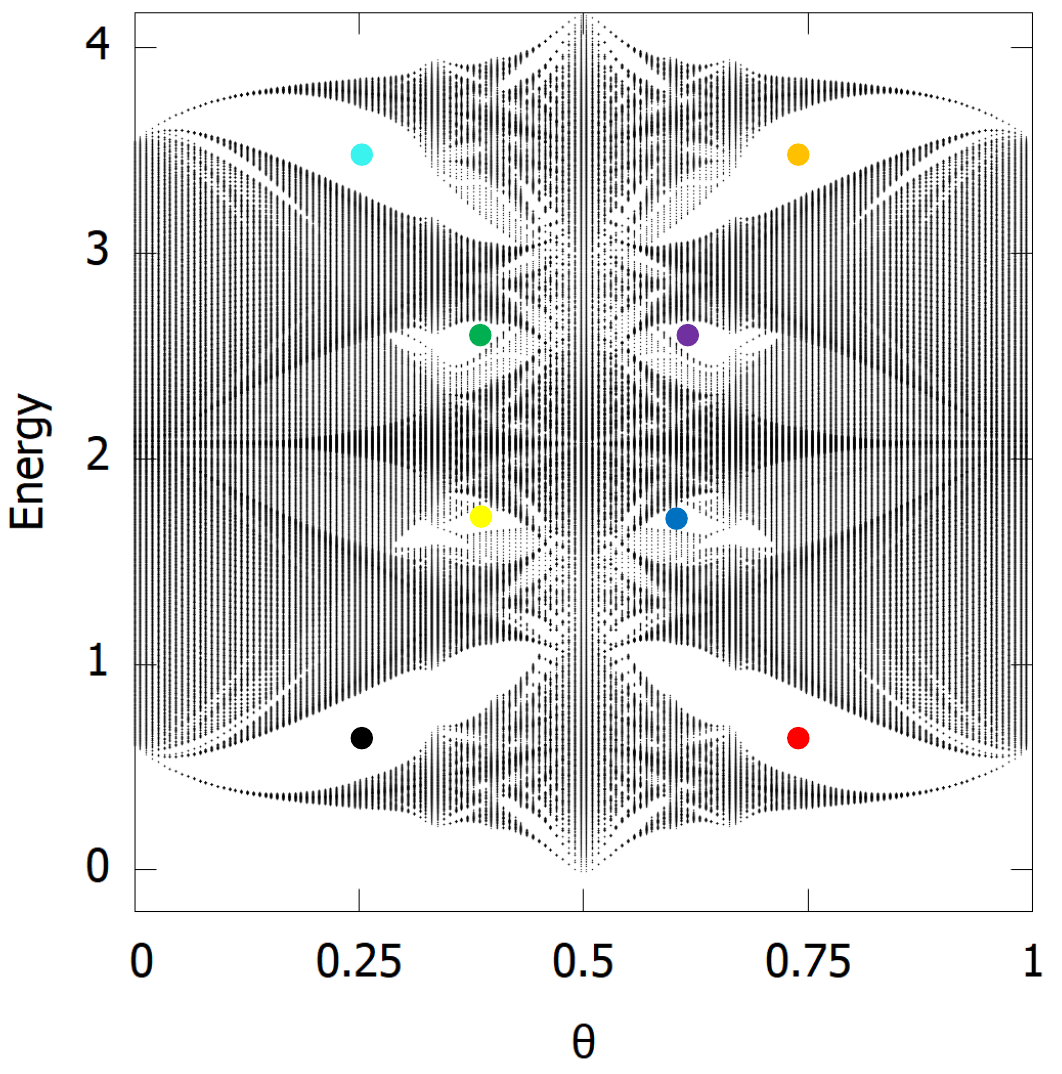}
	\caption{\small Bottom spectral island of the Hamiltonian~\eqref{Eq:GenHam} as a function of $\theta$, for the $M=2$ sector and $J_z=2.1$. The range of the parameter $\theta$ has been sampled at rational values $\theta_n = \frac{n}{|\Ll|}$ with $|\Ll|=151$. The spectral gaps have been labeled exactly as in Fig.~\ref{Fig:M2JZ0Spec}.}
	\label{Fig:M2JZ2p1ButterflyBottom}
\end{figure}

An immediate consequence of the simple form of the many-body potential in \eqref{Eq:IntM2} is that, in the presence of interaction, the algebra $\Aa_{\Theta_2}$ identified in section~\ref{Sec:M2NC} has been enhanced by precisely one projection and the correlated Hamiltonian belongs to the new algebra $C^\ast(\Aa_{\Theta_2},P_1)$. For reasons explained shortly, we are going to investigate first the corner sub-algebra
\begin{equation}
P_1 \, C^\ast(\Aa_{\Theta_2},P_1)\, P_1,
\end{equation}
and we start by identifying a few special elements. The first one is the unitary element
\begin{equation}
W_1 = \tfrac{2}{\alpha}\, P_1(U\otimes I)P_1 = \tfrac{2}{\alpha} \, P_1(I \otimes U)P_1 ,
\end{equation}
with $\alpha=1+e^{\imath 2 \pi \theta}$. More explicitly, $W_1$ is the unitary operator \cite{Footnote3}
\begin{equation}
W_1 = \sum_n e^{\imath 2\pi n \theta}|\psi_{n} \rangle \langle \psi_{n}|.
\end{equation}
Equivalently, $W_1$ can be defined as
\begin{equation}\label{Eq:W1}
W_1 = \tfrac{1}{\alpha}\, (U \otimes I + I \otimes U)P_1 = \tfrac{1}{\alpha} \, P_1(U \otimes I + I \otimes U).
\end{equation}
Let us point out that the projections of the following elementary operators cancel out:
\begin{equation} \label{Eq:Cancel1M2}
P_1(U^n \otimes U^m - U^m \otimes U^n)P_1 =0.
\end{equation}

\vspace{0.2cm}

The second element is
\begin{equation}
W_2 = P_1(T \otimes T)P_1 = P_1(T \otimes T) = (T \otimes T) P_1.
\end{equation}
Note that $W_2$ can be equivalently expressed as
\begin{equation}\label{Eq:W2}
W_2 = P_1(T \otimes T) = (T \otimes T) P_1,
\end{equation}
because $P_1$ and $T \otimes T$ commute. Furthermore, the projection of the following elementary operators cancel out:
\begin{equation}\label{Eq:Cancel2M2}
P_1(T^n \otimes T^m)P_1 =0, \quad m \neq n, 
\end{equation}
hence, they don't contribute with any useful elements to the corner sub-algebra. Now, by using Eqs.~\ref{Eq:W1} and \ref{Eq:W2}, it is straightforward to verify that
\begin{equation}
W_1 W_2 = e^{\imath 2 \pi \theta} W_2 W_1.
\end{equation}
The conclusion is that the non-commutative 2-torus is embedded in the corner sub-algebra
\begin{equation}
C^\ast(W_1,W_2) \hookrightarrow P_1 C^\ast(\Aa_{\Theta_2},P_1) P_1.
\end{equation} 

\vspace{0.2cm}

\begin{figure}[t!]
	\includegraphics[width=0.9\linewidth]{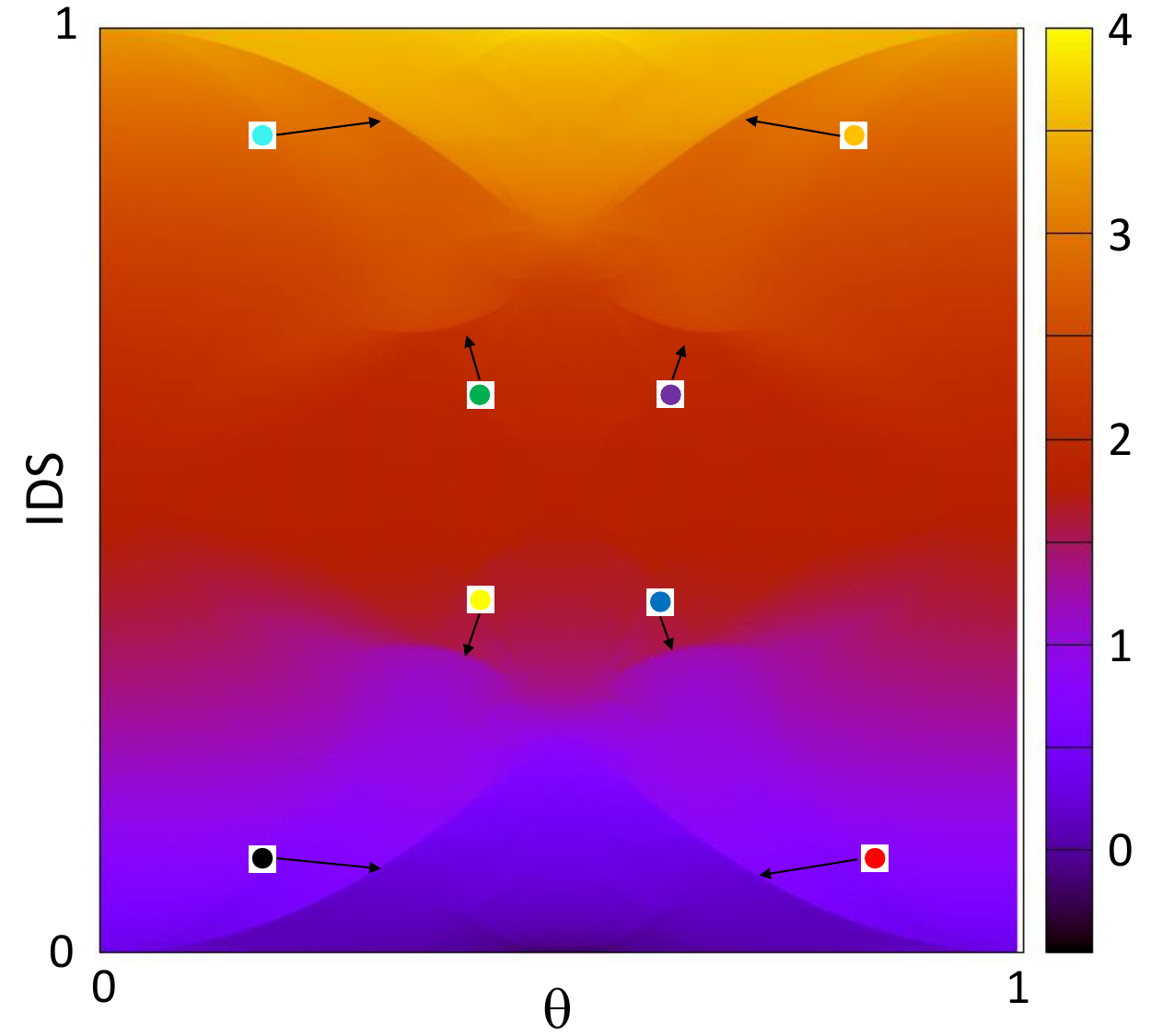}
	\caption{\small Numerical IDS as computed with Eq.~\eqref{Eq:IDSN1} using the spectrum reported in Fig.~\ref{Fig:M2JZ2p1ButterflyBottom}, as a function of $\theta$ and energy. The details of the simulation are the same as in Fig.~\ref{Fig:M2JZ2p1ButterflyBottom}. The features associtated to the abrupt changes in collors are correlated with the gaps marked in Fig.~\ref{Fig:M2JZ2p1ButterflyBottom}. They give the IDS values inside the spectral gaps, which can be fit exactly as in Fig.~\ref{Fig:GapLabels_M2JZ0}.}
	\label{Fig:M2JZ2p1IDSBottom}
\end{figure}   

We now establish the connection between the corner sub-algebra investigated above and the top spectral butterfly reported in Fig.~\ref{Fig:BulkSpecL151M2Jz2p1Eps0p0Top}. For this, let $P_{\rm Top}=I-P_{G_0}$ be the spectral projection onto the whole top island of the spectrum. Then, all the projections associated with the gaps seen in the spectrum reported in Fig.~\ref{Fig:BulkSpecL151M2Jz2p1Eps0p0Top} belong to the corner sub-algebra
\begin{equation}
P_{\rm Top}C^\ast(\Aa_{\Theta_2},P_1) P_{\rm Top}.
\end{equation} 
By re-scaling the Hamiltonian \eqref{Eq:GenHam} by $J_z$, one sees that, in the limit $J_z \rightarrow \infty$, the non-interacting part becomes a small perturbation and, as such,
\begin{equation}
P_{\rm Top} \rightarrow P_1 \ \ {\rm as} \ \  J_z \rightarrow \infty.
\end{equation}
This means that, for $J_z$ large enough, $\|P_{\rm Top} - P_1\| \leq 1$, in which case there exists a unitary operator $\Gamma \in C^\ast(\Aa_{\Theta_2},P_1)$  such that $P_{\rm Top} = \Gamma P_1 \Gamma^\ast$ \cite[p.~18]{ParkBook}. As a  consequence,
\begin{equation}
P_{\rm Top} \, C^\ast(\Aa_{\Theta_2},P_1) \, P_{\rm Top}  = (\Gamma P_1 \Gamma^\ast) C^\ast(\Aa_{\Theta_2},P_1) (\Gamma P_1 \Gamma^\ast).
\end{equation}
Since $\Gamma$ is a unitary operator from the algebra $C^\ast(\Aa_{\Theta_2},P_1)$, we automatically have
\begin{equation}
\Gamma^\ast \, C^\ast(\Aa_{\Theta_2},P_1) \, \Gamma = C^\ast(\Aa_{\Theta_2},P_1)
\end{equation}
as sets and algebras, and
\begin{equation}
P_{\rm Top} \, C^\ast(\Aa_{\Theta_2},P_1) \, P_{\rm Top} = \Gamma \, \Big ( P_1 C^\ast(\Aa_{\Theta_2},P_1) P_1 \Big ) \, \Gamma^\ast.
\end{equation}
The conclusion is that the corner sub-algebra $P_1 C^\ast(\Aa_{\Theta_2},P_1) P_1$ analyzed above and the sub-algebra $P_{\rm Top} \, C^\ast(\Aa_{\Theta_2},P_1) \, P_{\rm Top}$ which supplies the gap projection for the spectrum in Fig.~\ref{Fig:BulkSpecL151M2Jz2p1Eps0p0Top} are isomorphic.

\vspace{0.2cm}

\begin{figure*}[t!]
	\includegraphics[width=0.9\linewidth]{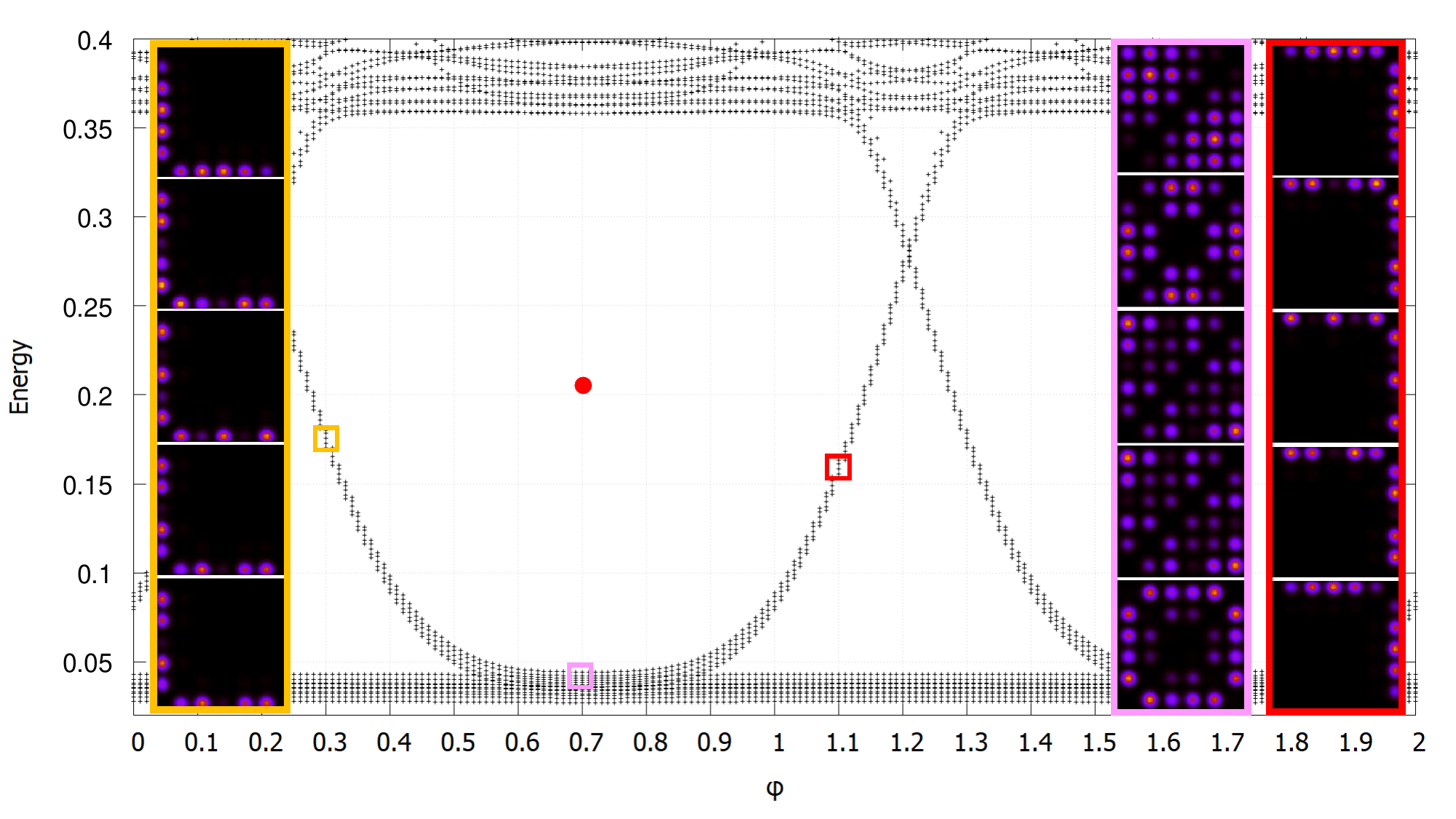}
	\caption{\small Visualization of the chiral edge modes associated to the bottom spectral island of the Hamiltonian~\eqref{Eq:GenHam} in the $M=2$ sector. The eigenvalues picked from different values of $\varphi$ and the corresponding wave-functions are color-coded. The simulation parameters are $r=0.45$, $\theta=\frac{2+\sqrt{5}}{5}$, $J_z=3$ and $|\Ll|=41$. The intensity maps represent the probabilities for two spins at locations $n$ and $m$ to be flipped along the chain. The colored dot label of the gap is correlated with the one in  Fig.~\ref{Fig:M2JZ2p1ButterflyBottom}.}
	\label{Fig:EdgeModeM2Jz3EM3}
\end{figure*}

At this point, we established that the non-commutative 2-torus sits inside $P_{\rm Top} \, C^\ast(\Aa_\Theta,P_1) \, P_{\rm Top}$ but is there anything else inside this algebra? The cancelations stated in Eqs.~\eqref{Eq:Cancel1M2} and \eqref{Eq:Cancel2M2} suggest that there is nothing else. For confirmation, we turn our attention on the IDS data reported in Fig.~\ref{Fig:GapLabels_M2JZ2p1}. To generate this plot, we used the spectra ${\rm Spec}$ from Fig.~\ref{Fig:BulkSpecL151M2Jz2p1Eps0p0Top} and the formula
\begin{equation}
{\rm IDS}(E) = \frac{\big |{\rm Spec} \cap [0,E] \big |}{|{\rm Spec} |}.
\end{equation}
The results in Fig.~\ref{Fig:GapLabels_M2JZ2p1} demonstrate that the IDS values inside every visible gap in Fig.~\ref{Fig:BulkSpecL151M2Jz2p1Eps0p0Top} can be explained by the K-theoretic predictions \eqref{Eq:IDS1Pred} derived from the non-commutative 2-torus. Furthermore, the K-Theoretic indices derived in Fig.~\ref{Fig:GapLabels_M2JZ2p1} are in perfect agreement with the count and the slopes of the chiral edge bands reported in Fig.~\ref{Fig:PhiDependenceM2Top}.  As such, we can state with confidence that the sub-algebra $P_{\rm Top} C^\ast(\Aa_{\Theta_2},P_1) P_{\rm Top}$ is in fact the non-commutative 2-torus.

\vspace{0.2cm}

Representations of the wave-functions corresponding to the chiral edge modes emerged in the top spectral island are supplied in Fig.~\ref{EdgeModeM2Jz3EM1}.  The intensity maps seen there render the probabilties $|\alpha_{n,m}|^2$ for two spins to be flipped at position $n$ and $m$ along the chain as functions of $n,m \in \{1,\ldots,\Ll\}$. Equivalently, $\alpha_{n,m}$ are the coefficients appearing in the expansion $|\Psi \rangle= \sum_{n,m} \alpha_{n,m} S_n^+ S_m^+|M= 0\rangle$ of the wave-functions. As expected, the two flipped spins always occupy neighboring sites and, as a consequence, the wave-functions are concentrated on two neighboring diagonals. Furthermore, when the eigenvalues are inside a bulk spectral gap, clear localizations at either the righ or left edges of the chain are observed, depending on the chirality of the bands.

\vspace{0.2cm}

A refined representation of the bottom spectral butterfly separated by the interaction and already identified in Fig.~\ref{Fig:JzSpecEvolution} is shown in Fig.~\ref{Fig:M2JZ2p1ButterflyBottom}. Its corresponding IDS map is reported in Figs.~\ref{Fig:M2JZ2p1IDSBottom}. The resemblance between this data and the one reported in Figs.~\ref{Fig:M2JZ0Spec} and \ref{Fig:GapLabels_M2JZ0} is very strong and it leaves little doubt that the spectral projections from the bottom spectral islands are generated from the non-commutative 4-torus. To confirm, we have verified that, indeed, the K-theoretic labels derived in Fig.~\ref{Fig:GapLabels_M2JZ0} apply identically to the IDS reported in Fig.~\ref{Fig:M2JZ2p1IDSBottom}. 

\vspace{0.2cm}

The spectral gaps in the bottom island of the spectrum remain topological and the bulk boundary principle stated in \eqref{Eq:BBM2} continues to apply. In Fig.~\ref{Fig:EdgeModeM2Jz3EM3}, we show the chiral modes that emerge in one of the large bulk spectral gaps from Fig~\ref{Fig:M2JZ2p1ButterflyBottom} when open boundary conditions are considered. As expected, we see not just one but a bundle of chiral bands. For the particular simulation in Fig.~\ref{Fig:EdgeModeM2Jz3EM3}, there are five bands in this bundle, but, in general, their will increase proportionally to $|\Ll|$, which is a consequence of \eqref{Eq:BBM2}. Fig.~\ref{Fig:EdgeModeM2Jz3EM3} also reports the profiles of the chiral modes and, in contrast to what we have seen for the top spectral island, this time only one of the flipped spins is localized at the edge of the chain and the other one is delocalized throughout the chain.

\begin{figure*}[t!]
	\includegraphics[width=0.9\linewidth]{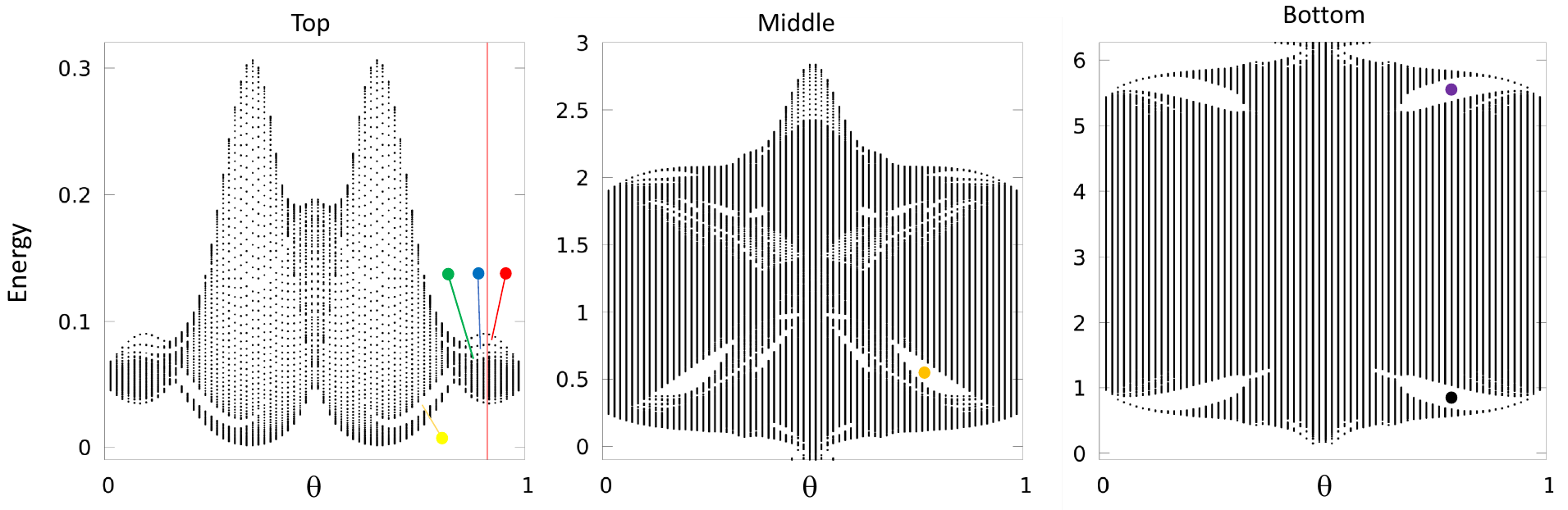}
	\caption{\small Top spectral island of Hamiltonian \eqref{Eq:GenHam} as a function of $\theta$, for $M=3$ and $J_z=4$. The simulation parameters are $r=0.45$ and $|\Ll|=71$. The red vertical line indicates the value $\theta=\frac{1+\sqrt{3}}{3}$, where the bulk-boundary correspondence is probed. Several spectral gaps have been labeled for future reference.}
	\label{Fig:BulkSpecL71M3Jz4TopMidBot}
\end{figure*}

\subsection{The $M=3$ sector}
\label{Sec:M3Corr}

\begin{figure*}[t!]
	\includegraphics[width=\linewidth]{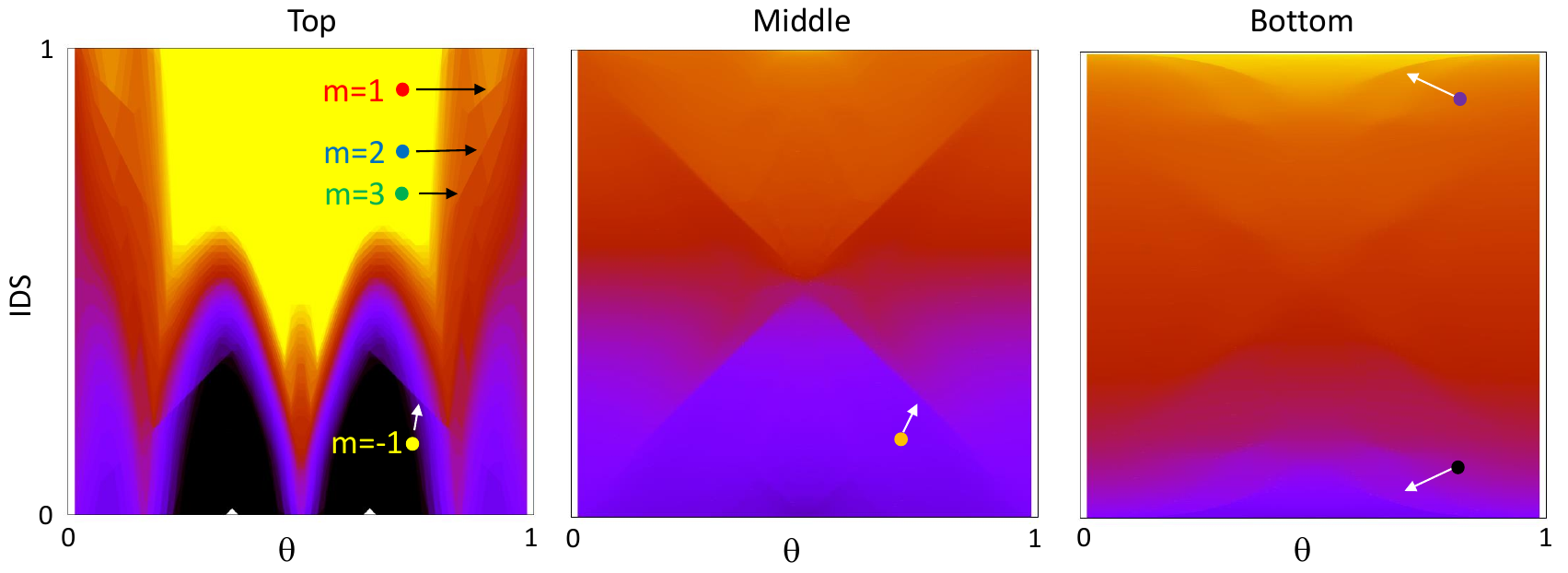}
	\caption{\small Numerical IDS maps of the top, middle and bottom spectra displayed in Fig.~\ref{Fig:BulkSpecL71M3Jz4TopMidBot}. The IDS values inside the spectral gaps marked in Fig.~\ref{Fig:BulkSpecL71M3Jz4TopMidBot} are indicated with arrows and colored dots. K-theoretic gap labels are displayed in the first panel.}
	\label{Fig:BulkIDSL71M3Jz4TopMidBot}
\end{figure*}

Refined representations of the three spectral islands observed in Fig.~\ref{Fig:JzSpecEvolution} for the $M=3$ sector are supplied in Fig.~\ref{Fig:BulkSpecL71M3Jz4TopMidBot} and their corresponding IDS maps are reported in Fig.~\ref{Fig:BulkIDSL71M3Jz4TopMidBot}. Clear straight lines can be identified in the IDS maps of the top and middle spectral butterflies, while the IDS map for the bottom spectral butterfly is identical to the one in Fig.~\ref{Fig:GapLabels_M3JZ0} for the un-correlated case. Furthermore, when the spectral islands are computed with open boundary conditions as a function of $\varphi$ and at fixed $\theta$, topological chiral modes emerge. Explaining and quantifying these empirical observations are the main goals of this section.

\vspace{0.2cm}

The interaction potential in the Hamiltonian \eqref{Eq:GenHam}, when restricted to the $M=3$ sector, reduces to
\begin{equation}\label{Eq:SpecD}
J_z \, \sum_{n} S^{z}_n S^z_{n+1} \mapsto \Lambda_1 P_1 + \Lambda_2 P_2+\Lambda_3 (I-P_1-P_2),
\end{equation}
where this time $P_1$ is the projection onto the subspace spanned by the states
\begin{equation}
|\psi_{n}\rangle = \tfrac{1}{\sqrt{3!}}\sum_{\rho}(-1)^\rho |n+\rho_1\rangle \otimes |n+\rho_2\rangle \otimes |n+\rho_3\rangle \in \Ff^{(-)}_3,
\end{equation}
with the sum running over the permutations $\rho$ of the set $\{0,1,2\}$, and $P_2$ is the projection onto the sub-space spanned by the states
\begin{equation}
|\psi_{n,k}\rangle = \tfrac{1}{\sqrt{3!}}\sum_{\rho}(-1)^\rho |n+\rho_1\rangle \otimes |n+\rho_2\rangle \otimes |n+\rho_3\rangle \in \Ff^{(-)}_3,
\end{equation}
with the sum running over the permutations $\rho$ of the set $\{0,1,k\}$ with $k\in \ZM \setminus \{-1,0,1,2\}$. When the energy is referenced from $\frac{|\Ll|}{4}J_z$, the above eigenvalues are
\begin{equation}
\Lambda_1 = -J_z, \ \  \Lambda_2 = -2J_z, \ \ \Lambda_3 = -3J_z.
\end{equation}
As one can see, the interaction potential becomes dominant for large $J_z$ and the eigenvalues $\Lambda_i$ start to separate from each other, which explains why the spectrum breaks into three islands at large $J_z$'s, as we have already seen in Fig.~\ref{Fig:JzSpecEvolution}.
 
\vspace{0.2cm}

An immediate consequence of the simple spectral decomposition \eqref{Eq:SpecD} is that, in the presence of interaction, the algebra $\Aa_{\Theta_3}$ identified in section~\ref{Sec:M3NC} has been enhanced by precisely two projections. Hence, the interacting Hamiltonian in the $M=3$ sector belongs to the algebra $C^\ast(\Aa_{\Theta_3},P_1,P_2)$. For reasons similar to the ones stated in Sec.~\ref{Sec:M2Corr}, we investigate first the corner sub-algebra
\begin{equation}
P_1 \, C^\ast(\Aa_{\Theta_3},P_1,P_2)\, P_1,
\end{equation}
and we start again by identifying a few special elements. The following relations identifies the first element $W_1$:
\begin{align}
P_1(U\otimes I \otimes I)P_1 & = P_1(I \otimes U \otimes I)P_1 \\ \nonumber
&= P_1(I \otimes I \otimes U)P_1 = \tfrac{\alpha}{3} \, W_1,
\end{align}
with $\alpha=1+e^{\imath 2 \pi \theta}+e^{\imath 4 \pi \theta}$. More explicitly, $W_1$ is the unitary operator
\begin{equation}
W_1 = \sum_n e^{\imath 2\pi n \theta}|\psi_{n} \rangle \langle \psi_{n}|.
\end{equation}
One can also verify that
\begin{align}\label{Eq:GoodW1RepM3}
W_1 & = \tfrac{1}{\alpha}(U \otimes I \otimes I + I \otimes U \otimes I + I \otimes I \otimes U)P_1 \\ \nonumber
& = \tfrac{1}{\alpha} P_1 (U \otimes I \otimes I + I \otimes U \otimes I + I \otimes I \otimes U).
\end{align} 
Note that any antisymmetric combinations of terms like $U^m \otimes U^n \otimes U^k$ cancel when sandwiched between $P_1$'s, hence such combinations do not contribute with elements in the corner algebra.

\vspace{0.2cm}

The second element is
\begin{equation}
W_2 = P_1(T \otimes T \otimes T)P_1.
\end{equation}
One can verify that $P_1$ commutes with $T\otimes T \otimes T$, hence $W_2$ can be also expressed as
\begin{equation}\label{Eq:GoodW2RepM3}
W_2 = P_1(T \otimes T\otimes T) = (T \otimes T \otimes T) P_1.
\end{equation}
Note that the projection of the following elementary operators cancel out:
\begin{equation}\label{Eq:CancelT}
P_1(T^n \otimes T^m \otimes T^k)P_1 = 0, 
\end{equation}
if $n$, $m$ $k$ are not all equal. Hence, they do not contribute with any useful elements for the corner algebra. 

\vspace{0.2cm}

Now, using the representations  \eqref{Eq:GoodW1RepM3} and \eqref{Eq:GoodW2RepM3} for $W_1$ and $W_2$, respectively, it is straightforward to verify that
\begin{equation}
W_1 W_2 = e^{\imath 2 \pi \theta} W_2 W_1.
\end{equation}
The conclusion is that the non-commutative 2-torus $\Aa_{\Theta_1}$ is embedded in the sub-algebra $P_1 C^\ast(\Aa_{\Theta_3},P_1,P_2) P_1$. As before, we denote by $P_{\rm Top}$ the spectral projection onto the top spectral island of the Hamiltonian. Since $P_{\rm Top} \rightarrow P_1$ in the limit $J_z \rightarrow \infty$, we can use the same arguments as in Sec.~\ref{Sec:M2Corr} to conclude that the corner algebra $P_{\rm Top} C^\ast(\Aa_{\Theta_3},P_1,P_2) P_{\rm Top}$ is isomorphic to $P_1 C^\ast(\Aa_{\Theta_3},P_1,P_2) P_1$. As such, the algebra $P_{\rm Top} C^\ast(\Aa_{\Theta_3},P_1,P_2) P_{\rm Top}$, which supply all spectral projections for the top island of the spectrum, contains a copy of the non-commutative 2-torus $\Aa_{\Theta_1}$.

\vspace{0.2cm}

\begin{figure*}[t]
	\includegraphics[width=\linewidth]{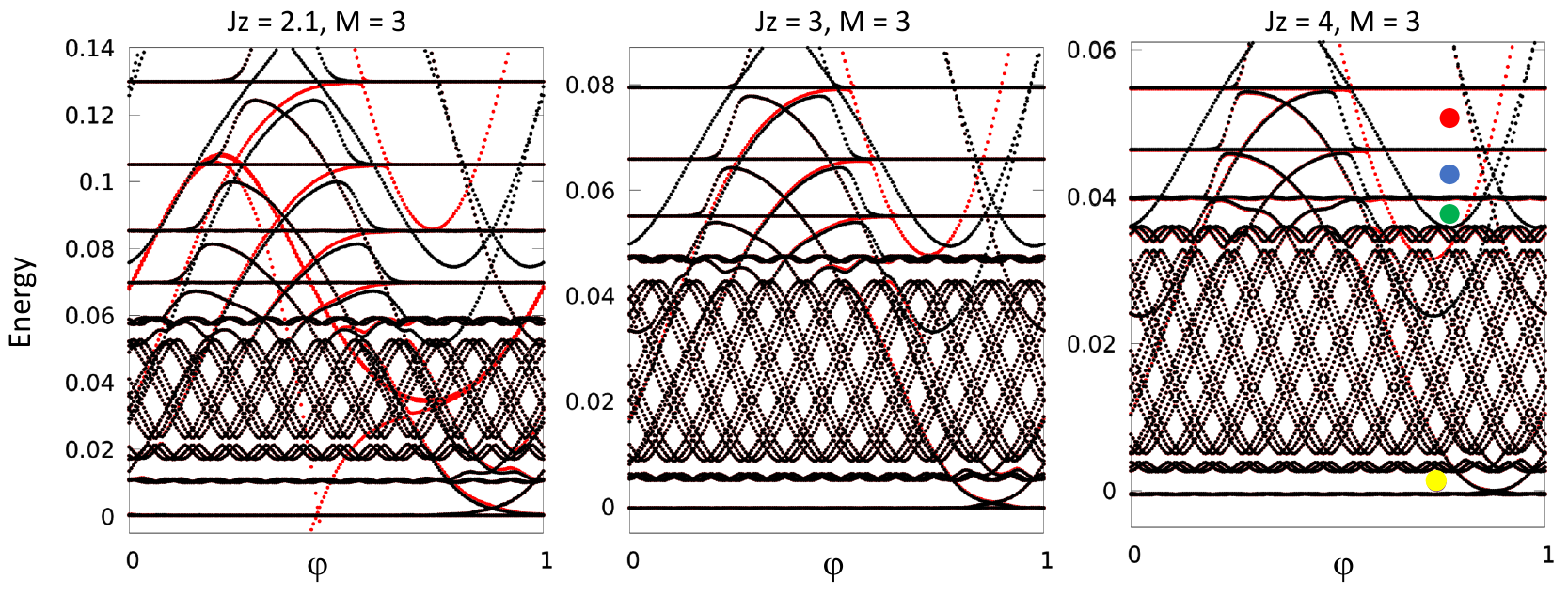}
	\caption{\small Top spectral island of Hamiltonian \eqref{Eq:GenHam} computed with open boundary condition as function of $\varphi$ at fixed $\theta=\frac{1 + \sqrt{3}}{3}$ (see vertical line in Fig.~\ref{Fig:BulkSpecL71M3Jz4TopMidBot}), and chain length $|\Ll|=41$. The colored dots labeling the gaps are correlated with the ones in  Fig.~\ref{Fig:BulkSpecL71M3Jz4TopMidBot}. The spectra were computed in two ways, with (red lines) and without (black lines) a defect potential on the left edge of the chain. This enabled us to identify the chiral bands located at the left edge of the chain, which are the ones for which the black and the red simulations do not overlap. Specifically, the gaps \textcolor{red}{\large $\bullet$}/\textcolor{blue}{\large $\bullet$}/\textcolor{green}{\large $\bullet$} display 1/2/3 positively sloped chiral bands localized at the left edge, respectively, while the \textcolor{yellow}{\large $\bullet$}-gap displays one negatively sloped chiral bands localized at the left edge. This is in perfect agreement with gap labels derived in Fig.~\ref{Fig:BulkIDSL71M3Jz4TopMidBot}.}
	\label{Fig:PhiDependenceM3Top}
\end{figure*}

The cancelations mentioned in Eq.~\eqref{Eq:CancelT} and the ones related to the $U$ operator suggest that $P_{\rm Top} C^\ast(\Aa_{\Theta_3},P_1,P_2) P_{\rm Top}$ is in fact identical to the 2-torus. This is further supported by the fact that all features identified in the IDS map in Fig.~\ref{Fig:BulkIDSL71M3Jz4TopMidBot} can be explained by the predictions from Eq.~\ref{Eq:IDS1Pred} based on the K-theory of the non-commutative 2-torus. Furthermore, the topological chiral bands emerging in the top spectrum when open boundary conditions are used, shown in Fig.~\ref{Fig:PhiDependenceM3Top}, are in perfect agreement with the gap labels derived from the IDS map in Fig.~\ref{Fig:BulkIDSL71M3Jz4TopMidBot}. 

\vspace{0.2cm}

\begin{figure*}[t!]
	\includegraphics[width=1\linewidth]{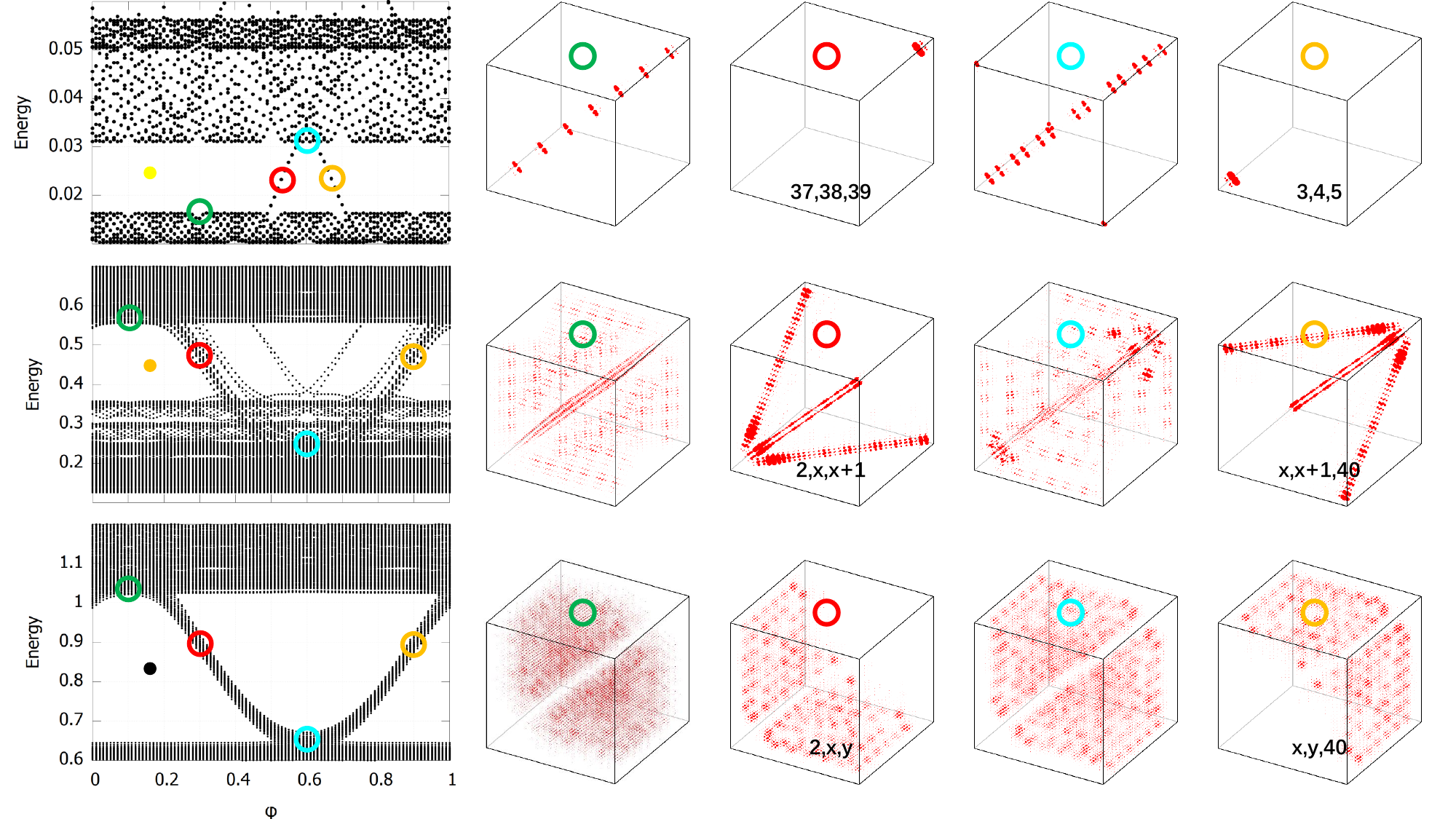}
	\caption{\small Visualization of the chiral edge modes associated to three spectral islands (top, middle, and bottom island from top to bottom, respectively) of the Hamiltonian~\eqref{Eq:GenHam} in the $M=3$ sector. The eigenvalues and the corresponding wave-functions are color-coded. The simulation parameters are $r=0.45$, $\theta=\frac{1+\sqrt{2}}{3}$, and $|\Ll|=41$. The value of $J_z$ is 4 for top and bottom islands, and 8 for middle one. The size of the points in the 3D renderings represents the probabilities for three spins at locations $n$, $m$ and $l$ along the chain to be flipped. The numbers seen in some of the panels represent the coordinates $(n,m,l)$ where the probabilities take significant values. The colored dots labeling the gaps are correlated with the ones in  Fig.~\ref{Fig:BulkSpecL71M3Jz4TopMidBot}.}
	\label{Fig:EdgeModeM3}
\end{figure*}

Representations of the top edge modes emerged in the top island of the spectrum under open boundary conditions are reported in the top row of Fig.~\ref{Fig:EdgeModeM3}. As expected, when the eigenvalues are located on the positively sloped chiral band occuring in the spectral \textcolor{yellow}{\large $\bullet$}-gap with index $m=-1$, all three flipped spins are localized on the right edge of the chain and quite the opposite when the eigenvalue is located on the negatively sloped chiral band. The transition between the two occurs through a delocalization of the mode when the eigenvalue dives in the bulk spectrum.

\vspace{0.2cm}

The IDS map for the middle spectral butterfly in Fig.~\ref{Fig:BulkIDSL71M3Jz4TopMidBot} shows that the dominant gaps seen in the middle panel of Fig.~\ref{Fig:BulkSpecL71M3Jz4TopMidBot} have linear IDS dependency on $\theta$, which is characteristic of the non-commutative 2-torus. In the following, we demonstrate that the corner algebra $P_{\rm Mid}C^\ast(\Aa_{\Theta_3},P_1,P_2) P_{\rm Mid}$, which supplies the spectral projections for the middle spectral butterfly, contains a copy of the non-commutative 2-torus $\Aa_{\Theta_1}$. Here, $P_{\rm Mid}$ is the spectral projector onto the full middle spectral island. Since $P_{\rm Mid}\rightarrow P_2$ in the limit $J_z \rightarrow \infty$, it is again enough to show that the corner algebra $P_2C^\ast(\Aa_{\Theta_3},P_1,P_2) P_2$ contains a copy of $\Aa_{\Theta_1}$. We will actually show that each sub-algebras $P_2(k)C^\ast(\Aa_{\Theta_3},P_1,P_2) P_2(k)$ contains a copy of $\Aa_{\Theta_1}$, where $P_2(k)$ is the projection onto the subspace spanned by $|\psi_{n,k}\rangle$, with $n \in \ZM$ and fixed $k$. For this, one observes that 
\begin{align}
& (U \otimes I \otimes I+I \otimes U \otimes I +I \otimes I \otimes U)|\psi_{n,k} \rangle \\ \nonumber
& \quad = (1+e^{\imath 2 \pi \theta}+e^{\imath 2 k \pi \theta}) \, e^{\imath 2 n \pi \theta}|\psi_{n,k}\rangle.
\end{align}
As such, $U \otimes I \otimes I+I \otimes U \otimes I +I \otimes I \otimes U$ is a diagonal operator in our standard basis, hence it commutes with $P_2(k)$ and we can define
\begin{equation}
W_1(k) = \frac{1}{\alpha_k} (U \otimes I \otimes I+I \otimes U \otimes I +I \otimes I \otimes U)P_2(k)
\end{equation}
with $\alpha_k = 1+e^{\imath 2 \pi \theta}+e^{\imath 2 k \pi \theta}$, which is a unitary element from $P_2(k)C^\ast(\Aa_{\Theta_3},P_1,P_2) P_2(k)$. Then, if we consider 
\begin{equation}
W_2(k) = P_2(k)(T \otimes T \otimes T)P_2(k),
\end{equation}
one can easily verify that they obey the commutation relations
\begin{equation}
W_1(k) W_2(k) = e^{\imath 2 \pi \theta} W_2(k) W_1(k),
\end{equation}
for all allowed $k$'s.

\vspace{0.2cm}

This finding may be the explanation for the existence of the dominant spectral gap with linear IDS dependency on $\theta$. However, further studies are needed to decided if the algebra associated to the midle states contain elements that are outside the non-commutative tori found above. The topological boundary spectrum and the associated modes emerged under open boundary conditions in the middle island of the spectrum are reported in the middle row of Fig.~\ref{Fig:EdgeModeM3}. As one can see, the \textcolor{orange}{\large $\bullet$}-gap with index $m=-1$ contains an entire bundle of chiral bands and the edge modes, whose eigenvalues are located on a positively sloped chiral band, have one flipped spin localized on the right edge of the chain while the other two flipped spins are constraint to neighbouring sites $(x,x+1)$ throughout the chain. The situation is quite the opposite when the eigenvalues are located on a negatively sloped chiral band. These findings are entirely consistent with the bulk-boundary correpsondence of the non-commutative 2-tori found above.

\vspace{0.2cm}

The structure of the gaps seen in the bottom spectral butterfly in Fig.~\ref{Fig:BulkSpecL71M3Jz4TopMidBot} as well as the features seen in the corresponding IDS map reported in Fig.~\ref{Fig:BulkIDSL71M3Jz4TopMidBot} are very similar to the ones for the uncorrelated case. In fact, we can confirm that the prediction \eqref{Eq:IDS3Pred} based on the K-theory of the non-commutative 3-torus explains all the features resolved in the numerical IDS map. The chiral bands and the corresponding modes emerged under open boundary conditions are reported in the last row of Fig.~\ref{Fig:EdgeModeM3}. As one can see, there is a thick bundle of chiral edge bands, which is consistent with the bulk-boundary principle stated in Eq.~\eqref{Eq:BBM3} saying that the count of the chiral modes should be proportional with $|\Ll|$. The corresponding wave-functions have one flipped spin localized at one edge of the chain and the remaining flipped spins are delocalized over the entire length of the chain.

\section{Conclusions}

Even though the algebras generating the interacting Hamiltonians were found to be non-commutative tori, the topological states identified in Sec.~\ref{Sec:TopoGaps2} are correlated and have no equivalent in the non-interacting case. This is the case because the generators of these algebras contain the spectral projections corresponding to the different islands of the spectrum and they are outside of the algebra that generate the non-interacting Hamiltonians. As such, it is impossible to generate the gap projection analyzed in Sec.~\ref{Sec:TopoGaps2} using the algebras analyzed in Sec.~\ref{Sec:TopoGaps1}.

Although we have only analyzed the $M=2$ and $M=3$ sectors in the strongly correlated regime, we can already conjecture what is going to happen for a generic magnetization sector $M=d$ with $d$ finite. While these predictions do not cover yet the case of a finite magnetization density, they can be of intereset for practical applications that, perhaps, can be implemented with cold atom systems. 

\vspace{0.2cm}

By extrapolating the cases analized in Sec.~\ref{Sec:M2Corr} and \ref{Sec:M3Corr}, we predict that, in such generic magnetization sector, the spectrum will split into $d$ islands for large enough $J_z$'s. The top island will always be characterized by a single non-commutative 2-torus whose generators can be computed explicitly. Under open boundary conditions, boundary modes will appear with energies inside the bulk gaps and these modes have $d$ flipped spins localized close to a boundary. These clusters of flipped spins can be adiabatically transferred from one edge of the chain to the other by simply changing the phason, more precisely, the shape of the underlying pattern. Hence, we have uncovered a simple Thouless pump where, by selecting the magnetization sector, one can transfer quantized amounts of magnetization between the edges of a system, as we have seen in Figs.~\ref{EdgeModeM2Jz3EM1} and \ref{Fig:EdgeModeM3}. As is the case with any Thouless pump, this processes will be robust against moderate disorder.

\vspace{0.2cm}

We conjecture that the bottom spectral island in a generic $M=d$ sector is characterized topologically by the non-commutative $d$-torus. The boundary modes will have a hybrid character with one flipped spin pinned at one boundary and the reset of the flipped spins delocalized throughout the chain.

\vspace{0.2cm}

We also conjecture that the intermediate islands will be all characterized by families of non-commutative 2-tori, whose generators can be computed explictly as it was already done in the present study. The boundary modes will have a hybrid character with one flipped spin pinned at one boundary and the rest of the flipped spins delocalized along the chain. The latter, however, will cluster into tight formations of $d-k+1$ of flipped spins for the $k$-th spectral island, as we have already seen for the middle spectral island in Fig.~\ref{Fig:EdgeModeM3}.

\vspace{0.2cm}

It remains a completely open question how to apply the K-theoretic ideas to the case of finite magnetization density, that is, when the conserved value of the magnetization grows linearly with the legth of the chain. The difficulty is that, in such situations, the algebras we already identified change as one takes the thermodynamic limit. As such, one needs to identify the correct relations between these algebras in order to resolve the limit. This will definitely be part of our future investigations. 

\acknowledgments{E.P. and Y.L. are supported by the NSF Grant No DMR-1823800. L.F.S. is supported by the NSF Grant No. DMR-1936006. E.P. acknowledges additional financial support from the W.M. Keck Foundation.}


\begin{thebibliography}{10}
\bibitem{Footnote1} The positively/negatively sloped bands are counted with $\pm$ signs, respectively, and only the bands localized at the left end of the chain are included in the count.
\bibitem{Footnote2} The size of the system was kept small to facilitate the count of the edge modes.
\bibitem{Footnote3} The unitarity is considered inside the corner algebra, where $P_1$ stands for the unit.
\end{thebibliography}
\end{document}